\documentclass[aps,pre,twocolumn,10pt,showpacs,superscriptaddress,nofootinbib]{revtex4-1}
\usepackage{amsmath,amssymb}
\usepackage{graphicx,color,colortbl}
\usepackage{rotating,array,tabularx,booktabs}
\newcolumntype{Y}{>{\centering\arraybackslash}X}

\newcommand{\dif}{\mathrm{d}}%
\newcommand{\Eins}{\mathbf{1}}%
\newcommand{\Null}{\mathbf{0}}%
\newcommand{\fdif}{\operatorname{\delta}}
\newcommand{\Fdif}[2]{\frac{\fdif\!#1}{\fdif\!#2}}
\newcommand{\ii}{\mathrm{i}}%
\newcommand{\id}{\operatorname{id}}%
\newcommand{\Laplace}{\boldsymbol{\triangle}}%
\newcommand{\Nabla}{\vec{\nabla}}%
\newcommand{\tdif}[2]{\frac{\dif#1}{\dif#2}}%
\newcommand{\pdif}[2]{\frac{\partial#1}{\partial#2}}%
\newcommand{\R}{\mathbb{R}}%
\newcommand{\C}{\mathbb{C}}%
\newcommand{\Tr}{\operatorname{Tr}}%
\newcommand{\norm}[1]{\lVert#1\rVert}%
\newcommand{\ie}{i.\,e.}%
\newcommand{\ZT}[1]{\textquotedblleft#1\textquotedblright}%
\newcommand{\rs}{\vec{r}\hskip1pt'}%
\newcommand{\rss}{\vec{r}\hskip1pt''}%
\newcommand{\HO}{\hat{H}}%
\newcommand{\Heff}{\hat{H}_{\mathrm{eff}}}%
\newcommand{\PP}{\text{\large$\operatorname{P}$}\!\!}%
\newcommand{\limeps}{\lim_{\epsilon\to 0^{+}}}%

\begin{document}
\title{Extended dynamical density functional theory for colloidal mixtures\\ with temperature gradients}

\author{Raphael Wittkowski}\author{Hartmut L{\"o}wen}
\affiliation{Institut f{\"u}r Theoretische Physik II, Weiche Materie,
Heinrich-Heine-Universit{\"a}t D{\"u}sseldorf, D-40225 D{\"u}sseldorf, Germany}

\author{Helmut R. Brand}
\affiliation{Theoretische Physik III, Universit{\"a}t Bayreuth, D-95540 Bayreuth, Germany}

\date{\today}

\begin{abstract}
In the past decade, classical dynamical density functional theory (DDFT) has been developed and widely applied to 
the Brownian dynamics of interacting colloidal particles. One of the possible derivation routes of DDFT
from the microscopic dynamics is via the Mori-Zwanzig-Forster projection operator technique with slowly varying variables
such as the one-particle density. Here, we use the projection operator approach to extend DDFT into various directions: 
first, we generalize DDFT toward mixtures of $n$ different species of spherical colloidal particles. 
We show that there are in general nontrivial cross-coupling terms between the concentration fields
and specify them explicitly for colloidal mixtures with pairwise hydrodynamic interactions.
Secondly, we treat the energy density as an additional slow variable and derive formal expressions
for an extended DDFT containing also the energy density. The latter approach can in principle be 
applied to colloidal dynamics in a nonzero temperature gradient.
For the case without hydrodynamic interactions the diffusion tensor is diagonal, while thermodiffusion 
-- the dissipative cross-coupling term between energy density and concentration -- is nonzero in this limit. 
With finite hydrodynamic interactions also cross-diffusion coefficients assume a finite value. 
We demonstrate that our results for the extended DDFT contain the transport coefficients in the hydrodynamic limit 
(long wavelengths, low frequencies) as a special case.  
\end{abstract}


\maketitle


\section{\label{sec:introduction}Introduction}
While classical density functional theory has become a quite popular tool to calculate static properties of inhomogeneous fluids 
\cite{Evans1979,Singh1991,Loewen1994a,WuL2007,TarazonaCMR2008}, 
its generalization to dynamical, \ie, time-dependent, properties is much less advanced.\footnote{For a recent review, 
see Ref.\ \cite{EmmerichEtAl2012}.} 
Most progress has been achieved for completely overdamped Brownian particles, which are realized as colloidal suspensions 
\cite{Pusey1991,Loewen2010}.
In a seminal paper of the year 1999, Marconi and Tarazona \cite{MarconiT1999,MarconiT2000} have derived a 
dynamical density functional theory (DDFT) from the Langevin equations describing the motion of the individual particles. 
The resulting DDFT equation corresponds to the field-theoretical model B for a single scalar order parameter, 
where the current is proportional to the functional density derivative of the equilibrium free-energy functional 
(generalized Fick's law). 
In 2004, Archer and Evans \cite{ArcherE2004} have used the stochastically equivalent Smoluchowski picture to rederive 
this DDFT equation. 
In 2009, Espa{\~n}ol and L\"owen \cite{EspanolL2009} have employed the 
Mori-Zwanzig-Forster projection operator technique (MZFT) \cite{Mori1965,ZwanzigM1965,Forster1974,Grabert1982,Forster1990,Dhont1996,Zwanzig2001}
as a third derivation route by using the one-particle density as the only slow variable of the system.

Subsequently, DDFT has been generalized toward binary mixtures \cite{Archer2005,RothRA2009,HuangEP2010,LichtnerAK2012}
and anisotropic particles \cite{BiervR2007,RexWL2007,WittkowskiL2011} as well as to the dynamics of freezing
\cite{vanTeeffelenLL2008,KahlL2009} and wetting \cite{ArcherRT2010}. 
Moreover, solvent-mediated hydrodynamic interactions between colloids, which are typically neglected in the modeling although they are  
important for actual colloidal samples, have been included into DDFT for the one-component case \cite{RexL2008,RexL2009,Rauscher2010}. 
More recent generalizations concern particle self-diffusion in complex environments \cite{ArcherHS2007,BiervRDvdS2008}, 
externally imposed flow fields \cite{RauscherDKP2007,BraderK2011}, colloidal sedimentation \cite{RoyallDSvB2007}, 
and \ZT{active} self-propelled particles \cite{WensinkL2008,WittkowskiL2011}.

In this paper, we follow the route via the MZFT in order to derive an extended DDFT (EDDFT), 
which goes beyond former DDFT in two respects.  
At first, we consider a multicomponent mixture of $n$ different species of spherical, \ie, isotropic, colloidal particles.
Hitherto, calculations for binary mixtures \cite{Archer2005,RothRA2009,HuangEP2010,LichtnerAK2012} assumed a diagonal
mobility matrix. Here, we show that there are in general nontrivial cross-coupling terms between the concentration fields. 
We specify these non-diagonal terms explicitly for colloidal mixtures with pairwise hydrodynamic interactions.
Therefore, we establish the basic dynamical equations to apply DDFT to the dynamics of multicomponent colloidal systems 
including their hydrodynamic interactions.
This constitutes a classic colloid problem, which has been explored intensely over several decades
by using mode-coupling-like techniques \cite{WagnerK1991,NaegeleD1998,Lionberger2002,KleinN2010}, 
computer simulations \cite{RexL2008b,MilinkovicPD2011}, and experiments \cite{KaplanYP1992,VissersWRLRIvB2011}.
Secondly, we treat the energy density as an additional slow variable and derive formal expressions
for an EDDFT containing also the energy density. This applies to situations, where a nonzero temperature gradient 
is imposed leading to thermodiffusion, which is also known as the Ludwig-Soret effect 
\cite{LifshitzP1981,KoehlerW2002,Piazza2004,Wiegand2004,PiazzaP2008,Wuerger2010}.
The derived equations also incorporate the reciprocal effect, which is the so-called Dufour effect \cite{Balescu1997}, 
where a concentration gradient causes energy transport.  

The MZFT is also the standard derivation route for mode-coupling theory (MCT) describing the dynamics of liquids. 
MCT has been applied both to molecular \cite{Goetze1991} and colloidal liquids \cite{DieterichP1979,HessK1983,CichockiH1987,SzamelL1991}
and is by now an pivotal theory for the glass transition. 
Here, we are using the same theoretical framework of the MZFT to derive an EDDFT. 
Furthermore, we are using this common basis to compare EDDFT and MCT. 
A possible connection between DDFT and MCT has already been pointed out by Archer \cite{Archer2006,Archer2009}.

There is also a close connection of the MZFT with several versions of the classical MCT close to phase transitions 
\cite{Kawasaki1970,Kawasaki1974,Kawasaki1975,HohenbergH1977}, which have been pioneered by Kawasaki. 
For example, for the isotropic-nematic phase transition in liquid crystal side-chain polymers a projection operator technique 
of MZFT-type \cite{zwanzig1974,BixonZ1978} has been used to construct a MCT \cite{BrandK1994,KawasakiB1994},
which was able to explain the experimental results obtained by two groups \cite{ReysDGMLBD1988,ReysDCKM1992,SigelSSL1993} 
as being due to a dynamic nonlinear coupling between order-parameter variations and the strain tensor. 

The use of MZFT to hydrodynamic condensed systems started with the work by Forster
for nematic liquid crystals \cite{Forster1974,Forster1974b} and has been 
applied in the following to a number of complex fluids with spontaneously broken
continuous symmetries \cite{Forster1990}, in particular to the superfluid phases of
$^{3}$He \cite{BrandDG1979,BrandP1982b} and, more recently, to uniaxial magnetic gels
\cite{BohliusBP2004}. These applications were based on a generalization
of the use of correlation functions in the hydrodynamic regime with applications to
simple fluids \cite{KadanoffM1963} and superfluid $^{4}$He \cite{HohenbergM1965}.
Thus, for all extensions of DDFT to more variables the hydrodynamic regime of long
wavelengths and low frequencies emerges for all condensed systems 
as a natural limit to check the results obtained.  
Conversely, the EDDFT can be used to investigate how far the range of hydrodynamic
considerations can be extended to larger values of frequencies and wave vectors.

The paper is organized as follows: in Sec.\ \ref{sec:MZF}, we summarize in detail the 
technical aspects of the projection operator technique we use in a coherent fashion.
In Sec.\ \ref{sec:collmix}, we present the results of the application of the MZFT 
to colloidal mixtures in detail including the energy density as a variable.
Finally, we summarize our results and present a perspective for future
generalizations of the present work in Sec.\ \ref{sec:conclu}.

\section{\label{sec:MZF}Mori-Zwanzig-Forster technique}
The MZFT \cite{Mori1965,ZwanzigM1965,Forster1974} is described in detail in several textbooks 
\cite{Grabert1982,Forster1990,Dhont1996,Zwanzig2001}. 
Further below, we comprehensively summarize the essential ideas that are relevant for this paper and adjust the notation to the problem at hand.

\subsection{General formalism}
For the purpose of this paper, it is most appropriate, but in general not necessary, to consider a grand-canonical ensemble of systems of $N$ particles. 
The total ensemble $\hat{\Gamma}_{t}$ with Hamiltonian $\HO(\hat{\Gamma}_{t})$\footnote{\label{fn:U}In order to keep the following expressions simple, 
the Hamiltonian $\HO(\hat{\Gamma}_{t})$ is assumed to be not explicitly time-dependent. Nevertheless, it is possible to consider also systems with 
a time-dependent external potential $U_{1}(\vec{r},t)$ in the framework of the MZFT as it is described here, if the external potential varies sufficiently 
slowly with time so that it is approximately constant on microscopic time scales.} involves as canonical variables the 
$6N$ coordinates $q_{i}(t)$ and momenta $p_{i}(t)$ of the $N$ particles.
It can be described by the total probability density $\hat{\rho}(t)\equiv\hat{\rho}(\hat{\Gamma}_{t})$, which is given by the solution of the 
\textit{Liouville-von Neumann equation}\footnote{In case of a classical system, the commutator $[X,Y]/(\ii\hbar)$ should be replaced by 
the Poisson brackets $\{X,Y\}$ for any variables $X$ and $Y$.} 
\begin{equation}
\dot{\hat{\rho}}=-\hat{\mathcal{L}}\hat{\rho}=-\frac{\ii}{\hbar}[\HO,\hat{\rho}] \;,\qquad \hat{\rho}(t)=e^{-\hat{\mathcal{L}}t}\hat{\rho}(0)
\label{eq:Liouville_rho}
\end{equation}
with the Liouvillian $\hat{\mathcal{L}}(\hat{\Gamma}_{t})$, 
the imaginary unit $\ii$, the reduced Planck constant $\hbar=h/(2\pi)$, 
and the commutator $[X,Y]=XY-YX$ of $X$ and $Y$. 
Alternatively, it is also possible to describe the same system in terms of only a few \textit{relevant} variables 
$\hat{a}_{i}(t)\equiv\hat{a}_{i}(\vec{r},t)\equiv\hat{a}_{i}(\hat{\Gamma}_{t};\vec{r})$\footnote{\label{fn:gsr}The abbreviating notation 
$X_{i}(t)\equiv X_{i}(\vec{r},t)\equiv X_{i}(\hat{\Gamma}_{t};\vec{r})$ 
is used for several symbols $X$ in this paper. Summation about an index $i$ implies also integration 
over $\vec{r}$ for such symbols: 
$X_{i}Y_{i}\equiv \sum_{i}\int\!\dif^{3}r\,X_{i}(\vec{r},t)Y_{i}(\vec{r},t)$. 
(Einstein's sum convention is used throughout this paper.)} with $i=1,\dotsc,n$, which we assume to be real-valued in the following.
The corresponding relevant ensemble $\Gamma_{t}$ is associated with the relevant probability density
$\rho(t)\equiv\rho(\hat{\Gamma}_{t})$.
Using the relevant probability density $\rho(t)$ and the grand-canonical trace $\Tr$, which is given for classical systems by 
\begin{equation}
\Tr=\sum^{\infty}_{N=0}\frac{e^{\beta\mu N}}{N!h^{3N}}\int_{\hat{\Gamma}_{t}}\!\!\!\!\dif\hat{\Gamma}_{t}
\end{equation}
with the inverse thermal energy $\beta=1/(k_{\mathrm{B}}T)$, Boltzmann constant $k_{\mathrm{B}}$, 
absolute temperature $T$, chemical potential $\mu$, and ensemble differential 
$\dif\hat{\Gamma}_{t}=\dif q_{1}\dif p_{1}\dotsb\dif q_{3N}\dif p_{3N}$, 
the time-dependent ensemble average 
\begin{equation}
\langle X(0) \rangle_{t}=\Tr(\rho(t)X(0))=\Tr(\rho(0)X(t))
\end{equation}
of an arbitrary variable $X(t)$ can be defined. 
The averaged relevant variables 
$a_{i}(t)\equiv a_{i}(\vec{r},t)$\footnote{Notice that $a_{i}(t)\equiv a_{i}(\vec{r},t)$ does not depend on 
the total ensemble $\hat{\Gamma}_{t}$, but the generalized summation rule defined in footnote \ref{fn:gsr} applies also here.}
are given by 
\begin{equation}
a_{i}(t)=\langle \hat{a}_{i}(0)\rangle_{t}=\Tr(\rho(t) \hat{a}_{i}(0))=\Tr(\hat{\rho}(t) \hat{a}_{i}(0)) \;.
\end{equation}
For a given thermodynamic functional like the Helmholtz free-energy functional $\mathcal{F}$, their thermodynamic conjugates 
$a^{\natural}_{i}(t)\equiv a^{\natural}_{i}(\vec{r},t)$ can be obtained by functional differentiation: 
\begin{equation}
a^{\natural}_{i}=\Fdif{\mathcal{F}}{a_{i}} \;.
\end{equation}
A possible representation for the Helmholtz free-energy functional is 
\begin{equation}
\mathcal{F}=\Tr\!\big(\rho\HO\big)+\frac{1}{\beta}\Tr\!\big(\rho\ln(\rho)\big) \,.
\end{equation}
The Helmholtz free-energy functional $\mathcal{F}[\vec{a}]$\footnote{This functional $\mathcal{F}[\vec{a}]=\mathcal{F}[a_{1},\dotsc,a_{n}]$ is also called \textit{density functional} in the context of DDFT.} 
depends functionally on the averaged relevant variables $a_{i}(t)$ and is related to the grand-canonical functional $\Omega[\vec{a}^{\natural}]$, 
that depends functionally on the thermodynamic conjugates $a^{\natural}_{i}(t)$, by the Legendre transformation 
\begin{equation}
\Omega[\vec{a}^{\natural}]=\mathcal{F}[\vec{a}]-a^{\natural}_{i}a_{i} \;.
\end{equation}
A map from the total ensemble $\hat{\Gamma}_{t}$ onto the relevant ensemble $\Gamma_{t}$ is constituted by a suitable   
projection operator $\hat{\mathcal{P}}_{t}=1-\hat{\mathcal{Q}}_{t}$. 
This projection operator can be written as \cite{Grabert1982}
\begin{equation}
\hat{\mathcal{P}}_{t}X=\Tr(\rho(t) X)+(\hat{a}_{i}-a_{i}(t))\Tr\!\bigg(\pdif{\rho(t)}{a_{i}(t)}X\bigg) \,.
\label{eq:P}
\end{equation}
It projects onto a space that is spanned by the linearly independent\footnote{Equivalently, one could also choose a linearly dependent set of 
relevant variables $\hat{a}_{i}(t)$ and construct a projector that maps onto the space spanned by these variables. 
In this case, the unity $\id$ is only indirectly taken into account as a basis element and the corresponding contribution 
-- the first term on the right-hand-side of Eq.\ \eqref{eq:P} -- has to be omitted (see, for example, Ref.\ \cite{Forster1990}).} 
variables $\hat{a}_{i}(t)$ and the unity $\id$.

The MZFT consists in the application of this operator in order to obtain transport equations for the relevant variables $\hat{a}_{i}(t)$, 
that are equivalent to the \textit{Liouville-von Neumann equations} 
\begin{equation}
\dot{\hat{a}}_{i}=\hat{\mathcal{L}}\hat{a}_{i}=\frac{\ii}{\hbar}[\HO,\hat{a}_{i}] \;,\qquad 
\hat{a}_{i}(t)=e^{\hat{\mathcal{L}}t}\hat{a}_{i}(0) \;,
\label{eq:Liouville_A}
\end{equation}
by projecting out all irrelevant variables. 
These transport equations are given, for example, in Ref.\ \cite{Grabert1982} in its general form.

The dynamics of the reduced relevant variables $\Delta \hat{a}_{i}(t)=\hat{a}_{i}(t)-a_{i}(t)$ is given by the 
(exact) \textit{generalized Langevin equations} \cite{Grabert1982,Forster1990}  
\begin{equation}
\begin{split}
\Delta \dot{\hat{a}}_{i}(t) &=\Omega_{ij}(t)\Delta \hat{a}_{j}(t) + \!\int^{t}_{0}\!\!\!\!\:\!\dif t'\, K_{ij}(t,t')\Delta \hat{a}_{j}(t') \\
&\quad\:\!+ \hat{F}_{i}(t)
\end{split}
\label{eq:GLG}
\end{equation}
with the frequency matrix 
\begin{equation}
\Omega_{ij}(t) = \pdif{}{a_{j}(t)}\Tr(\rho(t)\dot{\hat{a}}_{i}) \;,
\end{equation}
the memory matrix 
\begin{equation}
\begin{split}
K_{ij}(t,t') &= \Tr\!\bigg(\pdif{\rho(t')}{a_{j}(t')}\,\hat{\mathcal{L}}\,\hat{\mathcal{Q}}_{t'}\:\!\hat{\mathcal{G}}(t',t)\dot{\hat{a}}_{i}\bigg) \\
&\quad\:\!-\dot{a}_{k}(t')\Tr\!\bigg(\frac{\partial^{2}\!\rho(t')}{\partial a_{j}(t')\partial a_{k}(t')}\:\!
\hat{\mathcal{G}}(t',t)\dot{\hat{a}}_{i}\bigg) \,,
\end{split}
\end{equation}
and the noise
\begin{equation}
\hat{F}_{i}(t)=\hat{\mathcal{Q}}_{0}\:\!\hat{\mathcal{G}}(0,t)\dot{\hat{a}}_{i} \;.
\end{equation}
Here, $\hat{\mathcal{G}}(t',t)$ is the time-ordered exponential operator 
\begin{equation}
\hat{\mathcal{G}}(t',t)=\mathcal{T}_{-}\exp\!\bigg(\int^{t}_{t'}\!\!\!\!\:\!\dif t''\:\!\hat{\mathcal{L}}\:\!\hat{\mathcal{Q}}_{t''}\bigg) \,,
\end{equation}
where the time-ordering operator $\mathcal{T}_{-}$ orders operators from left to right as time increases.

Furthermore, the dynamics of the averaged relevant variables $a_{i}(t)$ is described by the 
\textit{averaged Langevin equations} \cite{Grabert1982}
\begin{equation}
\begin{split}%
\dot{a}_{i}(t)&=\Tr(\rho(t)\dot{\hat{a}}_{i}) + \!\int^{t}_{0}\!\!\!\!\:\!\dif t'\, 
\Tr\!\big(\rho(t')\hat{\mathcal{L}}\,\hat{\mathcal{Q}}_{t'}\:\!\hat{\mathcal{G}}(t',t)\dot{\hat{a}}_{i}\big) \\
&\quad\:\!+F_{i}(t)
\end{split}%
\label{eq:ALE}%
\end{equation}
with the averaged noise $F_{i}(t)=\Tr(\rho(0)\hat{F}_{i}(t))$.

The frequency matrix $\Omega_{ij}(t)$ takes the instantaneous reversible contributions to the dynamics of the relevant variables into account. 
In linearized form, it is an equal-time commutator of field operators. 
Whenever the chosen relevant variables $\hat{a}_{i}(t)$ have a definite time-reversal behavior, the frequency matrix vanishes. 
The memory matrix $K_{ij}(t,t')$, on the other hand, comprises the non-instantaneous reversible contributions and all dissipative contributions 
to the dynamics of the relevant variables. 
The important finding that the memory matrix can also include (non-instantaneous) reversible contributions was first shown by Forster 
\cite{Forster1974b,Forster1974,Forster1990}.

\subsection{Special generalized probability density}
The transport equations for the relevant variables and their correlation functions 
are given in this section for the specific case of the 
generalized grand canonical probability density 
\begin{equation}
\rho(t)=\frac{1}{\Xi(t)}\,e^{-\beta \Heff(t)} 
\label{eq:rho}%
\end{equation}
with the grand-canonical partition sum $\Xi(t)$ and the effective Hamiltonian
\begin{equation}
\Heff(t)=\HO-a^{\natural}_{i}(t)\hat{a}_{i} \;.
\end{equation}
For this particular choice, the projection operator \eqref{eq:P} is specified as the Robertson projector\footnote{For abbreviation, 
elements of inverse matrices are denoted as $\mathrm{M}^{-1}_{ij}\equiv(\mathrm{M}^{-1})_{ij}$ with an arbitrary matrix $\mathrm{M}$ 
in this paper.} \cite{Robertson1966,Grabert1982,EspanolL2009} 
\begin{equation}
\!\,\:\!\,\; \hat{\mathcal{P}}_{t}X=\Tr(\rho(t) X)+(\hat{a}_{i}-a_{i}(t))\,\chi^{-1}_{ij}(t)
\Tr\!\Big(\pdif{\rho(t)}{a^{\natural}_{j}(t)}X\Big) \!\!\!\!  
\end{equation}
with the symmetric non-equilibrium susceptibility matrix 
\begin{equation}
\begin{split}%
\chi_{ij}(t)=\Fdif{a_{i}(t)}{a^{\natural}_{j}(t)}&=
\beta \Tr\!\big(\rho(t)(\hat{a}_{i}-a_{i}(t)) \hat{\mathfrak{E}}_{t} (\hat{a}_{j}-a_{j}(t))\big) \,,
\end{split}%
\raisetag{2ex}%
\end{equation}
the derivative\footnote{This expression follows directly from the operator identity \cite{Grabert1982,Forster1990}
\begin{equation*}
\tdif{}{x}\:\!e^{A(x)}=\int^{1}_{0}\!\!\!\!\dif\lambda\,e^{\lambda A(x)}\tdif{A(x)}{x}\,e^{-\lambda A(x)}e^{A(x)} 
\end{equation*}
with a linear operator $A(x)$.}%
\begin{equation}
\pdif{\rho(t)}{a^{\natural}_{i}(t)}=\beta \,\hat{\mathfrak{E}}_{t} (\hat{a}_{i}-a_{i}(t))\:\!\rho(t) \;, 
\end{equation}
and the operator 
\begin{equation}
\hat{\mathfrak{E}}_{t}X = \int^{1}_{0}\!\!\!\!\dif\lambda\,e^{-\lambda\beta\Heff(t)} X e^{\lambda\beta\Heff(t)} \;. 
\end{equation}
This operator can be omitted for a classical system: $\hat{\mathfrak{E}}_{t}X=X$.

\subsubsection{Non-equilibrium dynamics}
The particular choice \eqref{eq:rho} of $\rho(t)$ leads to the \textit{exact transport equations} \cite{Grabert1982} 
\begin{equation}
\dot{a}_{i}(t)=-B_{ij}(t)a^{\natural}_{j}(t)-\!\int^{t}_{0}\!\!\!\!\:\!\dif t'\,R_{ij}(t,t')a^{\natural}_{j}(t')
\label{eq:TG}%
\end{equation}
with the antisymmetric drift matrix 
\begin{equation}
B_{ij}(t)=\frac{\ii}{\hbar}\Tr\!\big(\rho(t)\:\![\hat{a}_{i},\hat{a}_{j}]\big)=-B_{ji}(t)  
\label{eq:Bij}%
\end{equation}
and the retardation matrix 
\begin{equation}
R_{ij}(t,t')=\beta \Tr\!\Big(\rho(t')\big(\hat{\mathcal{Q}}_{t'}\:\!\hat{\mathcal{G}}(t',t)\dot{\hat{a}}_{i}\big)\!\:\!
\big(\hat{\mathfrak{E}}_{t'}\dot{\hat{a}}_{j}\big)\!\Big) \,. 
\label{eq:RM}%
\end{equation}
Notice that these transport equations are applicable also far from thermodynamic equilibrium.

\subsubsection{Equilibrium correlations}
In thermodynamic equilibrium, the transport equations for equilibrium time correlation functions 
(so-called \textit{Kubo functions} \cite{Forster1990})
\begin{equation}
C_{ij}(t)=\langle\Delta \hat{a}^{\mathrm{eq}}_{i}(t) | \Delta \hat{a}^{\mathrm{eq}}_{j}(0)\rangle_{\mathrm{eq}} 
\end{equation}
with the equilibrium fluctuations $\Delta \hat{a}^{\mathrm{eq}}_{i}(t)=\hat{a}_{i}(t)-a_{i}^{\mathrm{eq}}$ have no noise term.
Here, the letters \ZT{eq} denote equilibrium quantities and Mori's scalar product is given by 
\begin{equation}
\langle X | Y \rangle_{\mathrm{eq}} = \Tr\!\big(\rho^{\mathrm{eq}} X \:\!\hat{\mathfrak{E}}^{\mathrm{eq}} Y \big) 
\label{eq:MSP}%
\end{equation}
with the equilibrium probability density 
\begin{equation}
\rho^{\mathrm{eq}}=\frac{1}{\Xi^{\mathrm{eq}}}\,e^{-\beta\HO} 
\label{eq:rhoeq}%
\end{equation}
and the operator 
\begin{equation}
\hat{\mathfrak{E}}^{\mathrm{eq}}X = \int^{1}_{0}\!\!\!\!\dif\lambda\,e^{-\lambda\beta\HO} X\, e^{\lambda\beta\HO} \;. 
\end{equation}
On the basis of Mori's scalar product \eqref{eq:MSP}, the equilibrium average 
$\langle X\rangle_{\mathrm{eq}}=\langle X | \id \rangle_{\mathrm{eq}}=\Tr(\rho^{\mathrm{eq}} X)$ is defined.
In the following, we present transport equations for the time correlation functions $C_{ij}(t)$.
These transport equations are at first given in position-time space $(\vec{r},t)$ and later analyzed in the context of 
linear response theory in Fourier-Laplace space $(\vec{k},z)$.

\vskip3mm
\paragraph{Position-time space:}
In the linear regime near equilibrium, the dynamics of the time correlation functions $C_{ij}(t)$ 
can be derived from Eq.\ \eqref{eq:TG} by linearization. The resulting transport equations are given by  
\cite{Grabert1982,Forster1990}
\begin{equation}
\dot{C}_{ij}(t) =\Omega^{\mathrm{eq}}_{ik}\,C_{kj}(t) + \!\int^{t}_{0}\!\!\!\!\:\!\dif t'\, K^{\mathrm{eq}}_{ik}(t-t')C_{kj}(t') 
\label{eq:Cij}%
\end{equation}
with the equilibrium frequency matrix 
\begin{equation}
\Omega^{\mathrm{eq}}_{ij}=-B^{\mathrm{eq}}_{ik}\,\chi^{\mathrm{eq}-1}_{kj}
\end{equation}
and the equilibrium memory matrix 
\begin{equation}
K^{\mathrm{eq}}_{ij}(t)=-R^{\mathrm{eq}}_{ik}(t)\,\chi^{\mathrm{eq}-1}_{kj} \;. 
\end{equation}
These equilibrium matrices depend on the equilibrium drift matrix\footnote{While the first equality in Eq.\ \eqref{eq:Bijeq} follows directly from 
Eq.\ \eqref{eq:Bij}, the second equality holds only in the linear regime near equilibrium and can be derived by linearization of the first term on the 
right-hand-side of Eq.\ \eqref{eq:ALE} and comparison with the equivalent Eq.\ \eqref{eq:TG}.}  
\begin{equation}
B^{\mathrm{eq}}_{ij}=\frac{\ii}{\hbar}\langle[\hat{a}_{i},\hat{a}_{j}]\rangle_{\mathrm{eq}}
=-\beta \langle\Delta\dot{\hat{a}}^{\mathrm{eq}}_{i} | \Delta\hat{a}^{\mathrm{eq}}_{j}\rangle_{\mathrm{eq}} \;,
\label{eq:Bijeq}
\end{equation}
on the equilibrium retardation matrix\footnote{In Eq.\ \eqref{eq:Req}, a redundant $\hat{\mathcal{Q}}^{\mathrm{eq}}$ is often inserted 
directly in front of $\dot{\hat{a}}_{i}$ in order to obtain a more symmetric expression.} 
\begin{equation}
R^{\mathrm{eq}}_{ij}(t) = \beta \big\langle \hat{\mathcal{Q}}^{\mathrm{eq}}\hat{\mathcal{G}}^{\mathrm{eq}}(t)\dot{\hat{a}}_{i} \big| 
\dot{\hat{a}}_{j} \big\rangle_{\mathrm{eq}} \;,
\label{eq:Req}%
\end{equation}
and on the static equilibrium susceptibility matrix 
\begin{equation}
\chi^{\mathrm{eq}}_{ij} = \beta \langle\Delta\hat{a}^{\mathrm{eq}}_{i} | \Delta\hat{a}^{\mathrm{eq}}_{j}\rangle_{\mathrm{eq}} \;. 
\end{equation}
Here, the equilibrium projector $\hat{\mathcal{P}}^{\mathrm{eq}}=1-\hat{\mathcal{Q}}^{\mathrm{eq}}$ is given by 
\begin{equation}
\hat{\mathcal{P}}^{\mathrm{eq}}X = \langle X\rangle_{\mathrm{eq}} + \beta \Delta\hat{a}^{\mathrm{eq}}_{i}\,\chi^{\mathrm{eq}-1}_{ij}\:\!
\langle\Delta\hat{a}^{\mathrm{eq}}_{j} | X\rangle_{\mathrm{eq}}
\end{equation}
and the equilibrium exponential operator is 
\begin{equation}
\hat{\mathcal{G}}^{\mathrm{eq}}(t)=e^{\hat{\mathcal{L}}\:\!\hat{\mathcal{Q}}^{\mathrm{eq}} t} \,.
\end{equation}
Notice that the linearized transport equations \eqref{eq:Cij} can be used to determine the equilibrium frequency matrix $\Omega^{\mathrm{eq}}_{ij}$ 
and the equilibrium memory matrix $K^{\mathrm{eq}}_{ij}(t)$ exactly. 
It is thus possible to calculate these matrices by the evaluation of equilibrium correlation functions obtained from experiments or 
microscopic simulations.  

\vskip3mm
\paragraph{Fourier-Laplace space:}
In Fourier-Laplace space (see appendix \ref{app:FLT}) the dynamical equations \eqref{eq:Cij} obtain the simpler form \cite{Forster1990} 
\begin{equation}
\big(z\delta_{ik}-\Omega^{\mathrm{eq}}_{ik}-\widetilde{K}^{\mathrm{eq}}_{ik}(z)\big)\widetilde{C}_{kj}(z)
=C_{ij}(0)=\frac{\chi^{\mathrm{eq}}_{ij}}{\beta} 
\label{eq:CijFL}%
\end{equation}
with the obvious solution 
\begin{equation}
\widetilde{C}_{ij}(z) = \beta^{-1} \big(z\:\!\Eins-\Omega^{\mathrm{eq}}-\widetilde{K}^{\mathrm{eq}}(z)\big)^{-1}_{ik}\:\!
\chi^{\mathrm{eq}}_{kj} \;. 
\label{eq:CijFLLsg}%
\end{equation}
Notice that $\widetilde{C}_{ij}(z)$, $C_{ij}(0)$, $\Omega^{\mathrm{eq}}_{ij}$, $\widetilde{K}^{\mathrm{eq}}_{ij}(z)$, and $\chi^{\mathrm{eq}}_{ij}$ 
are given in Fourier space, although their wave-vector dependence is not denoted explicitly here. 
Furthermore, $X(t)$ denotes a time-dependent quantity, $\widetilde{X}(\omega)$ its Fourier transform, and $\widetilde{X}(z)$ its Laplace transform
in this paragraph.  

The Fourier transformed equilibrium frequency matrix $\Omega^{\mathrm{eq}}_{ij}$ and the Fourier-Laplace transformed equilibrium memory matrix $\widetilde{K}^{\mathrm{eq}}_{ij}(z)$ in Eqs.\ \eqref{eq:CijFL} and \eqref{eq:CijFLLsg} are given by 
\begin{equation}
\Omega^{\mathrm{eq}}_{ij}=-B^{\mathrm{eq}}_{ik}\,\chi^{\mathrm{eq}-1}_{kj} \;, \quad
\widetilde{K}^{\mathrm{eq}}_{ij}(z)=-\widetilde{R}^{\mathrm{eq}}_{ik}(z)\,\chi^{\mathrm{eq}-1}_{kj} 
\end{equation}
with the Fourier transformed equilibrium drift matrix 
\begin{equation}
B^{\mathrm{eq}}_{ij} = \frac{\ii}{\pi}\int_{\R}\!\!\hskip-0.5pt\dif\omega\,\widetilde{\chi}''_{ij}(\omega) \;, 
\end{equation}
the Fourier-Laplace transformed equilibrium retardation matrix\footnote{The expression \eqref{eq:RijFL} has been symmetrized 
by insertion of redundant operators $\hat{\mathcal{Q}}^{\mathrm{eq}}$.} 
\begin{equation}
\widetilde{R}^{\mathrm{eq}}_{ij}(z)=\beta \langle\Delta\dot{\hat{a}}^{\mathrm{eq}}_{i} 
| \hat{\mathcal{Q}}^{\mathrm{eq}}(z-\hat{\mathcal{L}}^{\mathrm{eq}}_{\hat{\mathcal{Q}}})^{-1}
\hat{\mathcal{Q}}^{\mathrm{eq}} | \Delta\dot{\hat{a}}^{\mathrm{eq}}_{j}\rangle_{\mathrm{eq}} 
\label{eq:RijFL}%
\end{equation}
with the equilibrium self-adjoined reduced Liouvillian 
$\hat{\mathcal{L}}^{\mathrm{eq}}_{\hat{\mathcal{Q}}}=\hat{\mathcal{Q}}^{\mathrm{eq}}\hat{\mathcal{L}}\hat{\mathcal{Q}}^{\mathrm{eq}}$ 
\cite{Goetze2009}, and the Fourier transformed static equilibrium susceptibility matrix 
\begin{equation}
\chi^{\mathrm{eq}}_{ij}=\limeps \widetilde{\chi}_{ij}(z)\big\rvert_{z=-\epsilon} 
=\frac{1}{\pi}\int_{\R}\!\!\hskip-0.5pt\dif\omega\,\frac{\widetilde{\chi}''_{ij}(\omega)}{\omega} \;.
\end{equation}
Here, the dynamic susceptibility matrix 
\begin{equation}
\widetilde{\chi}_{ij}(z)=\frac{1}{\pi}\int_{\R}\!\!\hskip-0.5pt\dif\omega\,\frac{\widetilde{\chi}''_{ij}(\omega)}{\omega+\ii z} 
\end{equation}
and the absorptive response function 
\begin{equation}
\chi''_{ij}(t-t')=\frac{1}{2\hbar}\langle[\hat{a}_{i}(t),\hat{a}_{j}(t')]\rangle_{\mathrm{eq}} 
\end{equation}
have been introduced. 
As usual in the context of linear response theory, the absorptive response function appears as a contribution in the 
complex response function
\begin{equation}
\widetilde{\chi}_{ij}(\omega) = \widetilde{\chi}'_{ij}(\omega) + \ii\, \widetilde{\chi}''_{ij}(\omega)
\end{equation}
with the reactive part $\widetilde{\chi}'_{ij}(\omega)$ and the absorptive part $\widetilde{\chi}''_{ij}(\omega)$, 
whose non-diagonal elements are not necessarily real-valued. 
The reactive response function $\widetilde{\chi}'_{ij}(\omega)$ and the absorptive response function $\widetilde{\chi}''_{ij}(\omega)$ 
are dependent and related to each other by the  
Kramers-Kronig (dispersion) relations
\begin{equation}
\begin{split}%
\widetilde{\chi}'_{ij}(\omega)&=\frac{1}{\pi}\;\PP\int_{\R}\!\!\hskip-0.5pt\dif\omega'\,
\frac{\widetilde{\chi}''_{ij}(\omega')}{\omega'-\omega} \;, \\
\widetilde{\chi}''_{ij}(\omega)&=-\frac{1}{\pi}\;\PP\int_{\R}\!\!\hskip-0.5pt\dif\omega'\,
\frac{\widetilde{\chi}'_{ij}(\omega')}{\omega'-\omega} \;.
\end{split}%
\end{equation}
Also the time correlation functions $C_{ij}(t)$ can be expressed in terms of the absorptive response function: 
\begin{equation}
\dot{C}_{ij}(t)=\frac{2}{\ii\beta} \:\!\chi''_{ij}(t) \;. 
\end{equation}
Hence, their Laplace transforms $\widetilde{C}_{ij}(z)$ are given by 
\begin{equation}
\widetilde{C}_{ij}(z)=\frac{1}{\ii\pi\beta}\int_{\R}\!\!\hskip-0.5pt\dif\omega\,\frac{\widetilde{\chi}''_{ij}(\omega)}{\omega(\omega+\ii z)} \;.
\end{equation}

\subsection{Slow variables}
If the relevant variables vary sufficiently slowly with time so that there is a clear separation of time scales between the slowly relaxing 
relevant variables and the fast relaxing irrelevant variables, the transport equations \eqref{eq:TG} and \eqref{eq:Cij} can be simplified 
by neglecting contributions of order $\mathcal{O}(\dot{\hat{a}}^{3}_{i})$.
Using the expansion \cite{Zwanzig2001}
\begin{equation}
e^{\hat{\mathcal{Q}}_{t}\hat{\mathcal{L}}t} = e^{\hat{\mathcal{L}}t} + \mathcal{O}(\dot{\hat{a}}_{i}) \;,
\end{equation}
the retardation matrix \eqref{eq:RM} can be approximated by \cite{Grabert1982}
\begin{equation}
\!\;\; R_{ij}(t,t')=\beta \Tr\!\Big(\rho(t) \big(e^{\hat{\mathcal{L}}(t-t')}\hat{\mathcal{Q}}_{t}\dot{\hat{a}}_{i}\big)\!\:\!
\big(\hat{\mathfrak{E}}_{t}\dot{\hat{a}}_{j}\big)\!\Big)\!\:\! + \mathcal{O}(\dot{\hat{a}}^{3}_{k}) . \!\!\!\!\!\!\! 
\label{eq:RijSV}%
\end{equation}

\subsubsection{\label{sec:MA}Non-equilibrium dynamics}
The approximation \eqref{eq:RijSV} results in the simplified \textit{transport equations for slow variables} \cite{Grabert1982}
\begin{equation}
\dot{a}_{i}(t)=-B_{ij}(t)a^{\natural}_{j}(t)-\beta D_{ij}(t)a^{\natural}_{j}(t) \;.
\label{eq:TGe}
\end{equation}
This Markovian approximation is also applicable far from thermodynamic equilibrium, but it is not appropriate, if effects related to 
\ZT{long time tails} (like the glass transition \cite{Goetze2009}) are investigated \cite{Grabert1982}. 
The transport coefficients are given by the drift matrix $B_{ij}(t)$  
and the mobility matrix $D_{ij}(t)$.
The mobility matrix is given by the Green-Kubo-type expression
\begin{equation}
\begin{split}
D_{ij}(t)=\!\int^{\infty}_{0}\!\!\!\!\!\!\!\:\!\dif t'\,
\Tr\!\Big(\rho(t)\big(e^{\hat{\mathcal{L}}t'}\!\hat{\mathcal{Q}}_{t}\dot{\hat{a}}_{i}\big)\!\:\!
\big(\hat{\mathfrak{E}}_{t}\dot{\hat{a}}_{j}\big)\!\Big) \,. 
\end{split}
\end{equation}
In the classical limit, this expression simplifies to \cite{Zwanzig2001,EspanolL2009} 
\begin{equation}
D_{ij}(t)=\!\int^{\infty}_{0}\!\!\!\!\!\!\!\:\!\dif t'\, \Tr\!\Big(\rho(t)(\hat{\mathcal{Q}}_{t}\dot{\hat{a}}_{j}) 
e^{\hat{\mathcal{L}}t'}\!(\hat{\mathcal{Q}}_{t}\dot{\hat{a}}_{i})\Big) \,, 
\label{eq:Dij}%
\end{equation}
where a redundant $\hat{\mathcal{Q}}_{t}$ has been inserted in front of $\dot{\hat{a}}_{j}$ in order to symmetrize the expression.
Further redundant operators $\hat{\mathcal{Q}}_{t}$ could be inserted in the exponential function in Eq.\ \eqref{eq:Dij} 
by replacing the Liouvillian $\hat{\mathcal{L}}$ by the self-adjoint reduced Liouvillian 
$\hat{\mathcal{L}}^{\hat{\mathcal{Q}}}_{t}=\hat{\mathcal{Q}}_{t}\hat{\mathcal{L}}\hat{\mathcal{Q}}_{t}$.
Notice that the transport equations \eqref{eq:TGe} in combination with the approximation \eqref{eq:Dij} are exact up to the third order 
in $\dot{\hat{a}}_{i}$.

\subsubsection{Equilibrium correlations}
With the same approximation, the transport equations \eqref{eq:Cij} for the equilibrium time correlation functions $C_{ij}(t)$ become 
\begin{equation}
\dot{C}_{ij}(t) =\Omega^{\mathrm{eq}}_{ik}\,C_{kj}(t) + \Gamma^{\mathrm{eq}}_{ik}\,C_{kj}(t) \;.
\label{eq:CijSV}%
\end{equation}
Here, we introduced the transport matrix 
\begin{equation}
\Gamma^{\mathrm{eq}}_{ij} = -\beta D^{\mathrm{eq}}_{ik}\,\chi^{\mathrm{eq}-1}_{kj}
\end{equation}
with the mobility matrix 
\begin{equation}
D^{\mathrm{eq}}_{ij}=\!\int^{\infty}_{0}\!\!\!\!\!\!\!\:\!\dif t'\,  \langle\Delta\dot{\hat{a}}^{\mathrm{eq}}_{i} 
| \hat{\mathcal{Q}}^{\mathrm{eq}} \:\!e^{\hat{\mathcal{L}}t'}\! \hat{\mathcal{Q}}^{\mathrm{eq}} | 
\Delta\dot{\hat{a}}^{\mathrm{eq}}_{j}\rangle_{\mathrm{eq}} \;. 
\end{equation}

\subsection{Conserved quantities}
An important example for slowly relaxing variables are local densities of conserved quantities.
The transport equations of such conserved quantities $\hat{a}_{i}(\vec{r},t)$ can be written as conservation laws
\begin{equation}
\dot{\hat{a}}_{i}+\Nabla_{\vec{r}}\!\cdot\!\hat{\vec{J}}^{(i)}=0
\label{eq:Acl}%
\end{equation}
with local currents $\hat{\vec{J}}^{(i)}(\vec{r},t)$ corresponding to $\hat{a}_{i}(\vec{r},t)$. 
Analogous conservation laws hold for the averaged variables $a_{i}(\vec{r},t)$ with the averaged local currents 
$\vec{J}^{(i)}(\vec{r},t)=\Tr(\rho(t)\hat{\vec{J}}^{(i)}(\vec{r},0))$: 
$\dot{a}_{i}+\Nabla_{\vec{r}}\!\cdot\!\vec{J}^{(i)}=0$.

\subsubsection{Non-equilibrium dynamics}
Since only classical systems with slow variables are considered in the following, dynamical equations for the time-evolution of the 
averaged relevant variables $a_{i}(\vec{r},t)$ can be derived from Eqs.\ \eqref{eq:ALE}, \eqref{eq:TGe}, \eqref{eq:Dij}, and 
\eqref{eq:Acl}. These are the general classical \textit{extended DDFT equations} 
\begin{equation}
\begin{split}%
\!\!\dot{a}_{i}(\vec{r},t) = &- \Nabla_{\vec{r}}\!\cdot\!\Tr\!\big(\rho(t)\hat{\vec{J}}^{(i)}(\vec{r},0)\big) \\
&+\sum^{n}_{j=1}\Nabla_{\vec{r}}\!\cdot\!\!\int_{\R^{3}}\!\!\!\!\!\:\!\dif^{3}r'\,\beta D^{(ij)}(\vec{r},\rs\!,t)
\Nabla_{\rs}a^{\natural}_{j}(\rs\!,t)\!\!
\end{split}%
\label{eq:DDFTg}%
\end{equation}
with the diffusion tensor
\begin{equation}
\begin{split}%
&D^{(ij)}_{kl}(\vec{r},\rs\!,t)=\\
&\qquad\!\int^{\infty}_{0}\!\!\!\!\!\!\dif t'\,\Tr\!\Big(\rho(t)\big(\hat{\mathcal{Q}}_{t}\hat{J}^{(j)}_{l}(\rs\!,0)\big)e^{\hat{\mathcal{L}}t'}\!
\big(\hat{\mathcal{Q}}_{t}\hat{J}^{(i)}_{k}(\vec{r},0)\big)\!\Big) \,.
\end{split}%
\label{eq:Dijkl}\raisetag{3em}%
\end{equation}
If the variables $\hat{a}_{i}(t)$ are real and have definite time-reversal signatures, one can show that $D^{(ij)}_{kl}(\vec{r},\rs\!,t)$ 
is symmetric \cite{Forster1990}: $D^{(ij)}_{kl}(\vec{r},\rs\!,t)=D^{(ji)}_{lk}(\vec{r},\rs\!,t)$. 
This statement is known as \textit{Onsager's principle} \cite{LandauL1996}.

\subsubsection{Equilibrium correlations}
The assumption of conserved quantities can also be used to rearrange the transport equations \eqref{eq:CijSV} for the equilibrium time correlation 
functions $C_{ij}(\vec{r},\rs\!,t)$ into 
\begin{equation}
\dot{C}_{ij}(\vec{r},\rs\!,t) = \sum^{n}_{k=1}\Nabla_{\vec{r}}\!\cdot\!\!\int_{\R^{3}}\!\!\!\!\!\:\!\dif^{3}r''\, 
L^{(ik)}_{\mathrm{eq}}(\vec{r},\rss)\:\! C_{kj}(\rss\!,\rs\!,t)
\label{eq:CijCQ}%
\end{equation}
with the total transport matrix 
\begin{equation}
L^{(ij)}_{\mathrm{eq}}(\vec{r},\rs) = \Omega^{(ij)}_{\mathrm{eq}}(\vec{r},\rs) + \Gamma^{(ij)}_{\mathrm{eq}}(\vec{r},\rs) \;. 
\end{equation}
This matrix includes the contributions 
\begin{equation}
\begin{split}%
\Omega^{(ij)}_{\mathrm{eq}}(\vec{r},\rs) &= -\sum^{n}_{k=1}\int_{\R^{3}}\!\!\!\!\!\:\!\dif^{3}r''\, 
B^{(ik)}_{\mathrm{eq}}(\vec{r},\rss)\,\chi^{\mathrm{eq}-1}_{kj}(\rss\!,\rs) 
\end{split}%
\raisetag{1em}%
\end{equation}
and 
\begin{equation}
\begin{split}%
\Gamma^{(ij)}_{\mathrm{eq}}(\vec{r},\rs) &= \sum^{n}_{k=1}\int_{\R^{3}}\!\!\!\!\!\:\!\dif^{3}r''\, 
\beta D^{(ik)}_{\mathrm{eq}}(\vec{r},\rss)\,\Nabla_{\rss}\chi^{\mathrm{eq}-1}_{kj}(\rss\!,\rs) 
\end{split}%
\raisetag{1em}%
\end{equation}
with the equilibrium drift tensor 
\begin{equation}
B^{(ij)}_{\mathrm{eq}}(\vec{r},\rs) = \beta \langle \hat{\vec{J}}^{(i)}(\vec{r},0) | \Delta\hat{a}^{\mathrm{eq}}_{j}(\rs\!,0) \rangle_{\mathrm{eq}}
\end{equation}
and the equilibrium diffusion tensor 
\begin{equation}
\begin{split}%
D^{(ij)}_{\mathrm{eq}}(\vec{r},\rs) &= \!\int^{\infty}_{0}\!\!\!\!\!\!\!\:\!\dif t'\,  \langle\hat{\vec{J}}^{(i)}(\vec{r},0) 
| \hat{\mathcal{Q}}^{\mathrm{eq}} \:\!e^{\hat{\mathcal{L}}t'}\! \hat{\mathcal{Q}}^{\mathrm{eq}} | 
\hat{\vec{J}}^{(j)}(\rs\!,0)\rangle_{\mathrm{eq}} \;. 
\end{split}%
\raisetag{0.55em}%
\end{equation}

\section{
\label{sec:collmix}
Colloidal mixtures}
We now consider a mixture of $N_{\mathrm{c}}=\sum^{n}_{i=1}N_{i}$ isotropic colloidal particles of $n$ different species, 
where $N_{i}$ is the total number of particles of species $i\in\{1,\dotsc,n\}$\footnote{Notice that the meaning of $n$ in this section 
is different from its meaning in the previous section.}.  
These $N_{\mathrm{c}}$ colloidal particles are suspended in a molecular solvent consisting of $N_{0}$ 
small isotropic particles (molecules) of the same type. 

The MZFT is now used to derive an EDDFT equation for mixtures of colloidal particles. 
When $\vec{r}^{(i)}_{k}(t)$ denotes the position, $\vec{p}^{(i)}_{k}(t)$ the momentum, and $m_{i}$ the mass of the $k$th particle of species $i$, 
where $i=0$ corresponds to the molecules of the molecular solvent and $i>0$ corresponds to the colloidal particles, 
the Hamiltonian of the system is given by 
\begin{equation}
\HO(\hat{\Gamma}_{t},t)=\sum^{n}_{i=0}\sum^{N_{i}}_{k=1} \HO^{(i)}_{k}(\hat{\Gamma}_{t},t)
\label{eq:H}%
\end{equation}
with 
\begin{equation}
\begin{split}%
&\HO^{(i)}_{k}(\hat{\Gamma}_{t},t)=\frac{\vec{p}^{(i)2}_{k}}{2m_{i}}+U^{(i)}_{1}(\vec{r}^{(i)}_{k}\!,t) \\
&\qquad +\frac{1}{2}\sum^{n}_{j=0}\sum^{N_{j}}_{l=1} (1-\delta_{kl}\delta_{ij}) U^{(ij)}_{2}(\vec{r}^{(i)}_{k}\!-\vec{r}^{(j)}_{l}) \;.
\end{split}%
\label{eq:Hki}
\end{equation}
$U^{(i)}_{1}(\vec{r}^{(i)}_{k}\!,t)$ is the external potential acting on the 
particles of species $i$,  
$U^{(ij)}_{2}(\vec{r}^{(i)}_{k}\!-\vec{r}^{(j)}_{l})$ is the pair-interaction potential 
for two particles of species $i$ and $j$, respectively, and $\hat{\Gamma}_{t}$ is the total ensemble
introduced in the beginning of Sec.\ \ref{sec:MZF}.
To assure that the MZFT as described in Sec.\ \ref{sec:MZF} is applicable, the external potential is assumed to vary sufficiently slowly with time 
(see footnote \ref{fn:U}).   
The Liouvillian $\hat{\mathcal{L}}(\hat{\Gamma}_{t},t)$ corresponding to the Hamiltonian \eqref{eq:H} of the considered system is 
\begin{equation}
\hat{\mathcal{L}}=\sum^{n}_{i=0}\sum^{N_{i}}_{k=1} \Big(\Nabla_{\vec{p}^{(i)}_{k}}\HO\Big)\!\cdot\!\Nabla_{\vec{r}^{(i)}_{k}}
-\Big(\Nabla_{\vec{r}^{(i)}_{k}}\HO\Big)\!\cdot\!\Nabla_{\vec{p}^{(i)}_{k}} \;.
\end{equation}

\subsection{Relevant variables}
As relevant variables $\hat{a}_{i}(\vec{r},t)$ of the colloidal mixture, we choose the $n$ concentrations
\begin{equation}
\hat{c}_{i}(\vec{r},t)=\sum^{N_{i}}_{k=1} \delta\big(\vec{r}-\vec{r}^{(i)}_{k}(t)\big) 
\end{equation}
with $i\in\{1,\dotsc,n\}$ and the energy density 
\begin{equation}
\hat{\varepsilon}(\vec{r},t)=\sum^{n}_{i=0}\sum^{N_{i}}_{k=1} \HO^{(i)}_{k}(\hat{\Gamma}_{t},t) \:\!\delta\big(\vec{r}-\vec{r}^{(i)}_{k}(t)\big) \;.
\end{equation}
Their averages are denoted as $c_{i}(\vec{r},t)=\Tr(\rho(0)\hat{c}_{i}(\vec{r},t))$ and 
$\varepsilon(\vec{r},t)=\Tr(\rho(0)\hat{\varepsilon}(\vec{r},t))$ in the following. 

By considering only the $n$ concentrations $\hat{a}_{1}(\vec{r},t)=\hat{c}_{1}(\vec{r},t),\dotsc,\hat{a}_{n}(\vec{r},t)=\hat{c}_{n}(\vec{r},t)$ 
and the energy density $\hat{a}_{n+1}(\vec{r},t)=\hat{\varepsilon}(\vec{r},t)$ as relevant variables, we assume that the momentum variables 
$\vec{p}^{(i)}_{k}(t)$ relax much faster to local thermodynamic equilibrium than the position variables $\vec{r}^{(i)}_{k}(t)$ so that the 
momentum density can be neglected as a further dynamic variable on the characteristic time scale of the concentrations and of the energy density.
By this choice of relevant variables, we further assume that the concentration $\hat{c}_{0}(\vec{r},t)$ of the molecular solvent 
relaxes much faster than the concentrations $\hat{c}_{i}(\vec{r},t)$, $i>0$, of the colloidal particles. 

The concentrations $\hat{c}_{i}(\vec{r},t)$ and the energy density $\hat{\varepsilon}(\vec{r},t)$ are even under parity and time reversal. 
Furthermore, they are locally conserved, if there are no sources and sinks of particles and energy in the system. 

The corresponding currents follow from the Liouville equations $\dot{\hat{a}}_{i}+\{\HO,\hat{a}_{i}\}=0$ [see Eq.\ \eqref{eq:Liouville_A}] 
by comparison with Eq.\ \eqref{eq:Acl}.\footnote{For the derivation of $\hat{\vec{J}}^{\varepsilon}(\vec{r},t)$, 
the equation \cite{Grabert1982} 
\begin{equation*}
\begin{split}%
&\delta(\vec{r}-\rs)-\delta(\vec{r}-\rss) \\[-0.5ex]
&\qquad\quad=-\Nabla_{\vec{r}}\!\cdot\!\Big(\!(\rs-\rss)\int^{1}_{0}\!\!\!\!\!\dif\lambda\,\delta\big(\vec{r}-\rs+\lambda(\rs-\rss)\big)\!\Big)
\end{split}%
\end{equation*}
is helpful.}  
They are the particle number current 
\begin{equation}
\hat{\vec{J}}^{c_{i}}(\vec{r},t)=\sum^{N_{i}}_{k=1} \frac{\vec{p}^{(i)}_{k}}{m_{i}} \,\delta\big(\vec{r}-\vec{r}^{(i)}_{k}\big) 
\label{eq:Jc}%
\end{equation}
and the energy current 
\begin{equation}
\begin{split}%
&\hat{\vec{J}}^{\varepsilon}(\vec{r},t)=\sum^{n}_{i=0}\sum^{N_{i}}_{k=1} \frac{\vec{p}^{(i)}_{k}}{m_{i}}\:\! \HO^{(i)}_{k} 
\:\!\delta\big(\vec{r}-\vec{r}^{(i)}_{k}\big) \\
&\quad\:\!-\frac{1}{4}\sum^{n}_{i,j=0}\!\!\!\!\!\!\sum^{N_{i}}_{\begin{subarray}{c}k=1\\(k,i)\neq(l,j)\end{subarray}}\!\!\!\!\!\sum^{N_{j}}_{l=1}
\Big(\Nabla_{\vec{r}^{(ij)}_{kl}}U^{(ij)}_{2}(\vec{r}^{(ij)}_{kl})\!\Big)\!\!\:\!\cdot\!\!\:\!
\Big(\frac{\vec{p}^{(i)}_{k}}{m_{i}}+\frac{\vec{p}^{(j)}_{l}}{m_{j}}\Big) \\
&\qquad\:\!\times\vec{r}^{(ij)}_{kl}\!\!\:\!\int^{1}_{0}\!\!\!\!\dif\lambda\,\:\!
\delta\big(\vec{r}-\vec{r}^{(i)}_{k}\!\:\!+\lambda\:\!\vec{r}^{(ij)}_{kl}\big) 
\end{split}\raisetag{3.9em}%
\label{eq:Je}%
\end{equation}
with the dyadic product $\otimes$ and the notation $\vec{r}^{(ij)}_{kl}=\vec{r}^{(i)}_{k}-\vec{r}^{(j)}_{l}$, 
where all $\vec{r}^{(i)}_{k}$, $\vec{r}^{(ij)}_{kl}$, $\vec{p}^{(i)}_{k}$, and $\HO^{(i)}_{k}$ in Eqs.\ \eqref{eq:Jc} and \eqref{eq:Je}  
are to be taken at time $t$.

Since $\hat{\vec{J}}^{c_{i}}(\vec{r},t)$ and $\hat{\vec{J}}^{\varepsilon}(\vec{r},t)$ are of odd order in the momentum $\vec{p}^{(i)}_{k}(t)$,  
the averages $\Tr(\rho(t)\hat{\vec{J}}^{c_{i}})=0$ and $\Tr(\rho(t)\hat{\vec{J}}^{\varepsilon})=0$ vanish. 
This leads to the important invariance properties $\hat{\mathcal{Q}}_{t}\hat{\vec{J}}^{c_{i}}=\hat{\vec{J}}^{c_{i}}$ and 
$\hat{\mathcal{Q}}_{t}\hat{\vec{J}}^{\varepsilon}=\hat{\vec{J}}^{\varepsilon}$ [see Eq.\ \eqref{eq:P}].

\subsection{Transport equations}
Since the concentration fields $c_{i}(\vec{r},t)$ and the energy density $\varepsilon(\vec{r},t)$ are locally conserved, 
the EDDFT equations \eqref{eq:DDFTg} and the corresponding transport equations \eqref{eq:CijCQ}, respectively, can be applied. 
Due to the invariance of $\hat{c}_{i}(\vec{r},t)$ and $\hat{\varepsilon}(\vec{r},t)$ under time-reversal, the frequency matrix 
and therefore also the first term on the right-hand-side of Eq.\ \eqref{eq:DDFTg} vanish. 

\subsubsection{Non-equilibrium dynamics}
Application of Eqs.\ \eqref{eq:DDFTg} results in the following \textit{extended DDFT equations for colloidal mixtures}: 
{\allowdisplaybreaks%
\begin{align}%
\begin{split}%
\dot{c}_{i}(\vec{r},t) &= \sum^{n}_{j=1}\Nabla_{\vec{r}}\!\cdot\!\!\int_{\R^{3}}\!\!\!\!\!\:\!\dif^{3}r'\,\beta D^{(ij)}(\vec{r},\rs\!,t)
\Nabla_{\rs}c^{\natural}_{j}(\rs\!,t) \\
&\quad\:\!+\Nabla_{\vec{r}}\!\cdot\!\!\int_{\R^{3}}\!\!\!\!\!\:\!\dif^{3}r'\,\beta D^{(i\varepsilon)}(\vec{r},\rs\!,t)
\Nabla_{\rs}\varepsilon^{\natural}(\rs\!,t) \;,
\end{split}\label{eq:EDDFTa}\\%
\begin{split}%
\dot{\varepsilon}(\vec{r},t) &= \sum^{n}_{j=1}\Nabla_{\vec{r}}\!\cdot\!\!\int_{\R^{3}}\!\!\!\!\!\:\!\dif^{3}r'\, 
\beta D^{(\varepsilon j)}(\vec{r},\rs\!,t) \Nabla_{\rs}c^{\natural}_{j}(\rs\!,t) \\
&\quad\:\!+\Nabla_{\vec{r}}\!\cdot\!\!\int_{\R^{3}}\!\!\!\!\!\:\!\dif^{3}r'\,\beta D^{(\varepsilon\varepsilon)}(\vec{r},\rs\!,t)
\Nabla_{\rs}\varepsilon^{\natural}(\rs\!,t) \;.
\end{split}\label{eq:EDDFTb}%
\end{align}}%
The diffusion tensors $D^{(ij)}(\vec{r},\rs\!,t)$, $D^{(i\varepsilon)}(\vec{r},\rs\!,t)=(D^{(\varepsilon i)}(\vec{r},\rs\!,t))^{\mathrm{T}}$, 
and $D^{(\varepsilon\varepsilon)}(\vec{r},\rs\!,t)$ in the EDDFT equations \eqref{eq:EDDFTa} and \eqref{eq:EDDFTb} are given by 
{\allowdisplaybreaks%
\begin{align}%
\begin{split}%
D^{(ij)}_{kl}(\vec{r},\rs\!,t)&=
\!\int^{\infty}_{0}\!\!\!\!\!\!\!\:\!\dif t'\,\Tr\!\Big(\rho(t) \hat{J}^{c_{i}}_{k}(\vec{r},t') \hat{J}^{c_{j}}_{l}(\rs\!,0)\!\Big) \,,
\end{split}\label{eq:DTMa}\\%
\begin{split}%
D^{(i\varepsilon)}_{kl}(\vec{r},\rs\!,t)&=
\!\int^{\infty}_{0}\!\!\!\!\!\!\!\:\!\dif t'\,\Tr\!\Big(\rho(t) \hat{J}^{c_{i}}_{k}(\vec{r},t') \hat{J}^{\varepsilon}_{l}(\rs\!,0)\!\Big) \,,
\end{split}\label{eq:DTMb}\\%
\begin{split}%
D^{(\varepsilon\varepsilon)}_{kl}(\vec{r},\rs\!,t)&=
\!\int^{\infty}_{0}\!\!\!\!\!\!\!\:\!\dif t'\,\Tr\!\Big(\rho(t) \hat{J}^{\varepsilon}_{k}(\vec{r},t') \hat{J}^{\varepsilon}_{l}(\rs\!,0)\!\Big) \,.
\end{split}\label{eq:DTMc}%
\end{align}}%
They are associated with particle diffusion ($D^{(ij)}$), (inverse) thermodiffusion ($D^{(i\varepsilon)}$: Ludwig-Soret effect, 
$D^{(\varepsilon i)}$: Dufour effect), and heat conduction ($D^{(\varepsilon\varepsilon)}$), respectively. 
The EDDFT equations \eqref{eq:EDDFTa} and \eqref{eq:EDDFTb} in combination with the diffusion tensors \eqref{eq:DTMa}-\eqref{eq:DTMc} 
constitute the main result of this paper.

\subsubsection{Equilibrium correlations}
If the transport equations \eqref{eq:CijCQ} are applied, one obtains dynamical equations for the time correlation functions 
{\allowdisplaybreaks%
\begin{align}%
\begin{split}%
C_{ij}(\vec{r},\rs\!,t) &= \langle \Delta\hat{c}^{\mathrm{eq}}_{i}(\vec{r},t) | 
\Delta\hat{c}^{\mathrm{eq}}_{j}(\rs\!,0) \rangle_{\mathrm{eq}} \;, 
\end{split}\label{eq:DCij}\\%
\begin{split}%
C_{i\varepsilon}(\vec{r},\rs\!,t) &= \langle \Delta\hat{c}^{\mathrm{eq}}_{i}(\vec{r},t) | 
\Delta\hat{\varepsilon}^{\mathrm{eq}}(\rs\!,0) \rangle_{\mathrm{eq}} \;, 
\end{split}\label{eq:DCie}\\%
\begin{split}%
C_{\varepsilon\varepsilon}(\vec{r},\rs\!,t) &= \langle \Delta\hat{\varepsilon}^{\mathrm{eq}}(\vec{r},t) | 
\Delta\hat{\varepsilon}^{\mathrm{eq}}(\rs\!,0) \rangle_{\mathrm{eq}} \;. 
\end{split}\label{eq:DCee}%
\end{align}}%
These dynamical equations are given by 
{\allowdisplaybreaks%
\begin{align}%
\begin{split}%
\dot{C}_{ij}(\vec{r},\rs\!,t) &= \sum^{n}_{k=1}\Nabla_{\vec{r}}\!\cdot\!\!\int_{\R^{3}}\!\!\!\!\!\:\!\dif^{3}r''\, 
L^{(ik)}_{\mathrm{eq}}(\vec{r},\rss)\:\! C_{kj}(\rss\!,\rs\!,t) \\ 
&\quad\:\! +\Nabla_{\vec{r}}\!\cdot\!\!\int_{\R^{3}}\!\!\!\!\!\:\!\dif^{3}r''\, 
L^{(i\varepsilon)}_{\mathrm{eq}}(\vec{r},\rss)\:\! C_{\varepsilon j}(\rss\!,\rs\!,t) \;, 
\end{split}\label{eq:TGCij}\raisetag{3.2em}\\%
\begin{split}%
\dot{C}_{i\varepsilon}(\vec{r},\rs\!,t) &= \sum^{n}_{k=1}\Nabla_{\vec{r}}\!\cdot\!\!\int_{\R^{3}}\!\!\!\!\!\:\!\dif^{3}r''\, 
L^{(ik)}_{\mathrm{eq}}(\vec{r},\rss)\:\! C_{k\varepsilon}(\rss\!,\rs\!,t) \\ 
&\quad\:\! +\Nabla_{\vec{r}}\!\cdot\!\!\int_{\R^{3}}\!\!\!\!\!\:\!\dif^{3}r''\, 
L^{(i\varepsilon)}_{\mathrm{eq}}(\vec{r},\rss)\:\! C_{\varepsilon\varepsilon}(\rss\!,\rs\!,t) \;, 
\end{split}\label{eq:TGCie}\raisetag{3.2em}\\%
\begin{split}%
\dot{C}_{\varepsilon\varepsilon}(\vec{r},\rs\!,t) &= \sum^{n}_{k=1}\Nabla_{\vec{r}}\!\cdot\!\!\int_{\R^{3}}\!\!\!\!\!\:\!\dif^{3}r''\, 
L^{(\varepsilon k)}_{\mathrm{eq}}(\vec{r},\rss)\:\! C_{k\varepsilon}(\rss\!,\rs\!,t) \\ 
&\quad\:\! +\Nabla_{\vec{r}}\!\cdot\!\!\int_{\R^{3}}\!\!\!\!\!\:\!\dif^{3}r''\, 
L^{(\varepsilon\varepsilon)}_{\mathrm{eq}}(\vec{r},\rss)\:\! C_{\varepsilon\varepsilon}(\rss\!,\rs\!,t)  
\end{split}\label{eq:TGCee}\raisetag{3.2em}%
\end{align}}%
with the total transport matrices
{\allowdisplaybreaks%
\begin{align}%
\begin{split}%
L^{(ij)}_{\mathrm{eq}}(\vec{r},\rs) = \Omega^{(ij)}_{\mathrm{eq}}(\vec{r},\rs) + \Gamma^{(ij)}_{\mathrm{eq}}(\vec{r},\rs) \;, 
\end{split}\\%
\begin{split}%
L^{(i\varepsilon)}_{\mathrm{eq}}(\vec{r},\rs) = \Omega^{(i\varepsilon)}_{\mathrm{eq}}(\vec{r},\rs) 
+ \Gamma^{(i\varepsilon)}_{\mathrm{eq}}(\vec{r},\rs) \;, 
\end{split}\\%
\begin{split}%
L^{(\varepsilon\varepsilon)}_{\mathrm{eq}}(\vec{r},\rs) = \Omega^{(\varepsilon\varepsilon)}_{\mathrm{eq}}(\vec{r},\rs) 
+ \Gamma^{(\varepsilon\varepsilon)}_{\mathrm{eq}}(\vec{r},\rs) 
\end{split}%
\end{align}}%
consisting of the contributions 
{\allowdisplaybreaks%
\begin{align}%
\begin{split}%
\Omega^{(ij)}_{\mathrm{eq}}(\vec{r},\rs) = &-\sum^{n}_{k=1}\int_{\R^{3}}\!\!\!\!\!\:\!\dif^{3}r''\, 
B^{(ik)}_{\mathrm{eq}}(\vec{r},\rss)\,\chi^{\mathrm{eq}-1}_{kj}(\rss\!,\rs) \\
&-\int_{\R^{3}}\!\!\!\!\!\:\!\dif^{3}r''\, 
B^{(i\varepsilon)}_{\mathrm{eq}}(\vec{r},\rss)\,\chi^{\mathrm{eq}-1}_{\varepsilon j}(\rss\!,\rs) \;, 
\end{split}\label{eq:OmegaijMa}\raisetag{3.2em}\\%
\begin{split}%
\Omega^{(i\varepsilon)}_{\mathrm{eq}}(\vec{r},\rs) = &-\sum^{n}_{k=1}\int_{\R^{3}}\!\!\!\!\!\:\!\dif^{3}r''\, 
B^{(ik)}_{\mathrm{eq}}(\vec{r},\rss)\,\chi^{\mathrm{eq}-1}_{k\varepsilon}(\rss\!,\rs) \\
&-\int_{\R^{3}}\!\!\!\!\!\:\!\dif^{3}r''\, 
B^{(i\varepsilon)}_{\mathrm{eq}}(\vec{r},\rss)\,\chi^{\mathrm{eq}-1}_{\varepsilon\varepsilon}(\rss\!,\rs) \;, 
\end{split}\label{eq:OmegaijMb}\raisetag{3.2em}\\%
\begin{split}%
\Omega^{(\varepsilon\varepsilon)}_{\mathrm{eq}}(\vec{r},\rs) = &-\sum^{n}_{k=1}\int_{\R^{3}}\!\!\!\!\!\:\!\dif^{3}r''\, 
B^{(\varepsilon k)}_{\mathrm{eq}}(\vec{r},\rss)\,\chi^{\mathrm{eq}-1}_{k\varepsilon}(\rss\!,\rs) \\
&-\int_{\R^{3}}\!\!\!\!\!\:\!\dif^{3}r''\, 
B^{(\varepsilon\varepsilon)}_{\mathrm{eq}}(\vec{r},\rss)\,\chi^{\mathrm{eq}-1}_{\varepsilon\varepsilon}(\rss\!,\rs) 
\end{split}\label{eq:OmegaijMc}\raisetag{3.2em}%
\end{align}}%
and 
{\allowdisplaybreaks%
\begin{align}%
\begin{split}%
\Gamma^{(ij)}_{\mathrm{eq}}(\vec{r},\rs) &= \sum^{n}_{k=1}\int_{\R^{3}}\!\!\!\!\!\:\!\dif^{3}r''\, 
\beta D^{(ik)}_{\mathrm{eq}}(\vec{r},\rss)\,\Nabla_{\rss}\chi^{\mathrm{eq}-1}_{kj}(\rss\!,\rs) \\
&\quad\:\!+\int_{\R^{3}}\!\!\!\!\!\:\!\dif^{3}r''\, 
\beta D^{(i\varepsilon)}_{\mathrm{eq}}(\vec{r},\rss)\,\Nabla_{\rss}\chi^{\mathrm{eq}-1}_{\varepsilon j}(\rss\!,\rs) \;, 
\end{split}\label{eq:GammaijMa}\raisetag{3.3em}\\%
\begin{split}%
\Gamma^{(i\varepsilon)}_{\mathrm{eq}}(\vec{r},\rs) &= \sum^{n}_{k=1}\int_{\R^{3}}\!\!\!\!\!\:\!\dif^{3}r''\, 
\beta D^{(ik)}_{\mathrm{eq}}(\vec{r},\rss)\,\Nabla_{\rss}\chi^{\mathrm{eq}-1}_{k\varepsilon}(\rss\!,\rs) \\
&\quad\:\!+\int_{\R^{3}}\!\!\!\!\!\:\!\dif^{3}r''\, 
\beta D^{(i\varepsilon)}_{\mathrm{eq}}(\vec{r},\rss)\,\Nabla_{\rss}\chi^{\mathrm{eq}-1}_{\varepsilon\varepsilon}(\rss\!,\rs) \;, 
\end{split}\label{eq:GammaijMb}\raisetag{3.3em}\\%
\begin{split}%
\Gamma^{(\varepsilon\varepsilon)}_{\mathrm{eq}}(\vec{r},\rs) &= \sum^{n}_{k=1}\int_{\R^{3}}\!\!\!\!\!\:\!\dif^{3}r''\, 
\beta D^{(\varepsilon k)}_{\mathrm{eq}}(\vec{r},\rss)\,\Nabla_{\rss}\chi^{\mathrm{eq}-1}_{k\varepsilon}(\rss\!,\rs) \\
&\quad\:\!+\int_{\R^{3}}\!\!\!\!\!\:\!\dif^{3}r''\, 
\beta D^{(\varepsilon\varepsilon)}_{\mathrm{eq}}(\vec{r},\rss)\,\Nabla_{\rss}\chi^{\mathrm{eq}-1}_{\varepsilon\varepsilon}(\rss\!,\rs) \;.  
\end{split}\label{eq:GammaijMc}\raisetag{3.3em}%
\end{align}}%
Equations \eqref{eq:OmegaijMa}-\eqref{eq:GammaijMc} in turn depend on 
the equilibrium drift tensors 
{\allowdisplaybreaks%
\begin{align}%
\begin{split}%
B^{(ij)}_{\mathrm{eq}}(\vec{r},\rs) = \beta \langle \hat{\vec{J}}^{c_{i}}(\vec{r},0) | 
\Delta\hat{c}^{\mathrm{eq}}_{j}(\rs\!,0) \rangle_{\mathrm{eq}} \;, 
\end{split}\\%
\begin{split}%
B^{(i\varepsilon)}_{\mathrm{eq}}(\vec{r},\rs) = \beta \langle \hat{\vec{J}}^{c_{i}}(\vec{r},0) | 
\Delta\hat{\varepsilon}^{\mathrm{eq}}(\rs\!,0) \rangle_{\mathrm{eq}} \;, 
\end{split}\\%
\begin{split}%
B^{(\varepsilon\varepsilon)}_{\mathrm{eq}}(\vec{r},\rs) = \beta \langle \hat{\vec{J}}^{\varepsilon}(\vec{r},0) | 
\Delta\hat{\varepsilon}^{\mathrm{eq}}(\rs\!,0) \rangle_{\mathrm{eq}} \;, 
\end{split}%
\end{align}}%
on the equilibrium diffusion tensors 
{\allowdisplaybreaks%
\begin{align}%
\begin{split}%
D^{(ij)}_{\mathrm{eq}}(\vec{r},\rs) &= \!\int^{\infty}_{0}\!\!\!\!\!\!\!\:\!\dif t'\,  \langle\hat{\vec{J}}^{c_{i}}(\vec{r},0) 
| \hat{\vec{J}}^{c_{j}}(\rs\!,t')\rangle_{\mathrm{eq}} \;, 
\end{split}\\%
\begin{split}%
D^{(i\varepsilon)}_{\mathrm{eq}}(\vec{r},\rs) &= \!\int^{\infty}_{0}\!\!\!\!\!\!\!\:\!\dif t'\,  \langle\hat{\vec{J}}^{c_{i}}(\vec{r},0) 
| \hat{\vec{J}}^{\varepsilon}(\rs\!,t')\rangle_{\mathrm{eq}} \;, 
\end{split}\\%
\begin{split}%
D^{(\varepsilon\varepsilon)}_{\mathrm{eq}}(\vec{r},\rs) &= \!\int^{\infty}_{0}\!\!\!\!\!\!\!\:\!\dif t'\,  \langle\hat{\vec{J}}^{\varepsilon}(\vec{r},0) 
| \hat{\vec{J}}^{\varepsilon}(\rs\!,t')\rangle_{\mathrm{eq}} \;, 
\end{split}%
\end{align}}%
and on the static equilibrium susceptibility matrices 
{\allowdisplaybreaks%
\begin{align}%
\begin{split}%
\chi^{\mathrm{eq}}_{ij}(\vec{r},\rs) = \beta \langle\Delta\hat{c}^{\mathrm{eq}}_{i}(\vec{r},0) | 
\Delta\hat{c}^{\mathrm{eq}}_{j}(\rs\!,0)\rangle_{\mathrm{eq}} \;, 
\end{split}\\%
\begin{split}%
\chi^{\mathrm{eq}}_{i\varepsilon}(\vec{r},\rs) = \beta \langle\Delta\hat{c}^{\mathrm{eq}}_{i}(\vec{r},0) | 
\Delta\hat{\varepsilon}^{\mathrm{eq}}(\rs\!,0)\rangle_{\mathrm{eq}} \;, 
\end{split}\\%
\begin{split}%
\chi^{\mathrm{eq}}_{\varepsilon\varepsilon}(\vec{r},\rs) = \beta \langle\Delta\hat{\varepsilon}^{\mathrm{eq}}(\vec{r},0) | 
\Delta\hat{\varepsilon}^{\mathrm{eq}}(\rs\!,0)\rangle_{\mathrm{eq}} 
\end{split}%
\end{align}}%
with the equilibrium fluctuations $\Delta\hat{c}^{\mathrm{eq}}_{i}(\vec{r},t)=\hat{c}_{i}(\vec{r},t)-c_{i}^{\mathrm{eq}}(\vec{r})$ 
and $\Delta\hat{\varepsilon}^{\mathrm{eq}}(\vec{r},t)=\hat{\varepsilon}(\vec{r},t)-\varepsilon^{\mathrm{eq}}(\vec{r})$.

\subsection{Approximation of the diffusion tensors}
For an application of the EDDFT equations \eqref{eq:EDDFTa} and \eqref{eq:EDDFTb} to a particular system, suitable expressions for 
the diffusion tensors \eqref{eq:DTMa}-\eqref{eq:DTMc} are needed. 
A possibility to determine these diffusion tensors is the implementation of particle-resolved computer simulations \cite{Satoh2010}. 
Alternatively, analytical approximations for the diffusion tensors $D^{(ij)}(\vec{r},\rs\!,t)$, $D^{(i\varepsilon)}(\vec{r},\rs\!,t)$, 
and $D^{(\varepsilon\varepsilon)}(\vec{r},\rs\!,t)$ can be applied. 
Such approximate expressions are given in the following.

\subsubsection{No hydrodynamic interactions}
As first approximation, it is assumed that the considered system is sufficiently close to local thermodynamic equilibrium so that the 
relevant probability density $\rho(t)$ can be approximated by the equilibrium probability density $\rho^{\mathrm{eq}}$ 
[see Eq.\ \eqref{eq:rhoeq}] in Eqs.\ \eqref{eq:DTMa}-\eqref{eq:DTMc}.   
Secondly, we assume that the position variables relax much more slowly to local thermodynamic equilibrium than the momentum variables 
and that the external potential is approximately constant on microscopic length scales. 
Thirdly, we suppose that the position and momentum variables are statistically independent. 
Furthermore, the considered suspension shall be sufficiently dilute so that hydrodynamic interactions between the colloidal particles 
can be neglected and the momenta of different particles are uncorrelated.
Finally, we assume orientational isotropy for the momentum variables, \ie, 
$\langle\vec{p}\otimes\!\:\!\vec{p}\rangle_{\mathrm{eq}}=\frac{1}{3}\Eins
\langle\vec{p}\!\cdot\!\vec{p}\rangle_{\mathrm{eq}}$ and neglect the pair-interaction potential 
$U^{(ij)}_{2}(\vec{r}^{(i)}_{k}\!-\vec{r}^{(j)}_{l})$ in Eqs.\ \eqref{eq:Hki} and \eqref{eq:Je}.

With these assumptions, the diffusion tensors \eqref{eq:DTMa}-\eqref{eq:DTMc} can be approximated by
{\allowdisplaybreaks%
\begin{align}%
\begin{split}%
D^{(ij)}_{\mathrm{NH}}(\vec{r},\rs\!,t)&=D^{(i)}_{0}\:\!\Eins\,\delta_{ij}\delta(\vec{r}-\rs)c_{i}(\vec{r},t) \;,
\end{split}\label{eq:DI}\\[1.3ex]%
\begin{split}%
D^{(i\varepsilon)}_{\mathrm{NH}}(\vec{r},\rs\!,t)&=\kappa^{(i)}_{\mathrm{S}}\:\!\Eins\,\delta(\vec{r}-\rs)c_{i}(\vec{r},t) \;,
\end{split}\label{eq:DII}\\%
\begin{split}%
D^{(\varepsilon\varepsilon)}_{\mathrm{NH}}(\vec{r},\rs\!,t)&=\sum^{n}_{i=0} \kappa^{(i)}_{\mathrm{H}}\:\!\Eins\,\delta(\vec{r}-\rs)c_{i}(\vec{r},t) 
\end{split}\label{eq:DIII}%
\end{align}}%
with the transport coefficients
{\allowdisplaybreaks%
\begin{align}%
\begin{split}%
D^{(i)}_{0}=\frac{1}{3}\!\int^{\infty}_{0}\!\!\!\!\!\!\!\:\!\dif t'\,\langle\vec{v}_{i}(t')\!\cdot\!\vec{v}_{i}(0)\rangle_{\mathrm{eq}} \;,
\end{split}\label{eq:Di}\\%
\begin{split}%
\kappa^{(i)}_{\mathrm{S}}=\frac{1}{3}\!\int^{\infty}_{0}\!\!\!\!\!\!\!\:\!\dif t'\,\langle\vec{v}_{i}(t')\!\cdot\!\vec{v}_{i}(0)
\HO_{i}(0)\rangle_{\mathrm{eq}} \;,
\end{split}\label{eq:kappaS}\\%
\begin{split}%
\kappa^{(i)}_{\mathrm{H}}=\frac{1}{3}\!\int^{\infty}_{0}\!\!\!\!\!\!\!\:\!\dif t'\,\langle\vec{v}_{i}(t')\!\cdot\!\vec{v}_{i}(0)
\HO_{i}(t')\HO_{i}(0)\rangle_{\mathrm{eq}} 
\end{split}\label{eq:kappaH}%
\end{align}}%
where $\vec{v}_{i}(t)=\vec{p}^{(i)}(t)/m_{i}$ is the velocity of a colloidal particle of species $i$  
and $\HO_{i}(t)$ is its energy. 
These coefficients are associated with particle diffusion, thermodiffusion, and heat conduction, respectively. 
Notice that the diffusion tensor \eqref{eq:DI} is diagonal and that all diffusive cross-couplings in Eq.\ \eqref{eq:EDDFTa} vanish, 
if there are no hydrodynamic interactions between the colloidal particles.

\subsubsection{Hydrodynamic interactions}
A better approximation for the diffusion tensor \eqref{eq:DTMa}, that takes also diffusive cross-couplings into account, 
can be derived, if hydrodynamic interactions between the colloidal particles are taken into account.  
In order to do so, the derivation of the DDFT equation for a one-component suspension of colloidal particles with hydrodynamic interactions 
in Refs.\ \cite{RexL2008,RexL2009} is generalized and compared with Eq.\ \eqref{eq:EDDFTa}.   
This derivation starts from the Smoluchowski equation \cite{Dhont1996}
\begin{equation}
\dot{P}(\vec{r}^{N}\!,t)+\sum^{n}_{i=1}\sum^{N_{i}}_{k=1}\Nabla_{\vec{r}^{(i)}_{k}}\!\!\cdot\!\vec{J}^{(i)}_{\mathrm{P},k}(\vec{r}^{N}\!,t) = 0
\end{equation}
with the $N$-particle probability density $P(\vec{r}^{N}\!,t)$, where 
$\vec{r}^{N}=(\vec{r}^{(1)}_{1}\!,\dotsc,\vec{r}^{(n)}_{N_{n}})$ are the positions of all particles, 
the probability currents 
\begin{equation}
\vec{J}^{(i)}_{\mathrm{P},k}(\vec{r}^{N}\!,t)= 
-\sum^{n}_{j=1}\sum^{N_{j}}_{l=1}\mathrm{D}^{(ij)}_{kl}(\vec{r}^{N}) \vec{f}^{(j)}_{l}(\vec{r}^{N}\!,t) \;, 
\end{equation}
and the force densities
\begin{equation}
\vec{f}^{(j)}_{l}(\vec{r}^{N}\!,t)=
\Nabla_{\vec{r}^{(j)}_{l}}P(\vec{r}^{N}\!,t)+P(\vec{r}^{N}\!,t)\Nabla_{\vec{r}^{(j)}_{l}}\big(\beta U(\vec{r}^{N}\!,t)\big) \;. 
\end{equation}
Here, $\mathrm{D}^{(ij)}_{kl}(\vec{r}^{N})$ is a short-time diffusion tensor and $U(\vec{r}^{N}\!,t)$ denotes the total potential energy 
of the system. 
If the considered suspension is not too dense so that the particle distances are sufficiently large, the hydrodynamic interactions can be 
approximated on the two-particle level and higher-order hydrodynamic interactions are negligible. 

In case of only hydrodynamic pair-interactions, the short-time diffusion tensors $\mathrm{D}^{(ij)}_{kl}(\vec{r}^{N})$ 
can be written in the exact form \cite{HappelB1991,Dhont1996}
\begin{equation}
\begin{split}
&\mathrm{D}^{(ij)}_{kl}(\vec{r}^{N})=D^{(i)}_{0}\delta_{ij}\delta_{kl}\:\!\Eins \\
&\;\quad+ D^{(i)}_{0}\delta_{ij}\delta_{kl}\sum^{n}_{q=1}\sum^{N_{q}}_{p=1} (1-\delta_{kp}\delta_{iq}) 
\,\mathrm{h}^{(iq)}_{\mathrm{s}}(\vec{r}^{(iq)}_{kp}) \\
&\;\quad+ (1-\delta_{ij}\delta_{kl})D^{(j)}_{0} \:\!\mathrm{h}^{(ij)}_{\mathrm{c}}(\vec{r}^{(ij)}_{kl}) 
\end{split}
\label{eq:Dklij}%
\end{equation}
with the self- and cross-interaction functions 
\begin{equation}
\mathrm{h}^{(ij)}_{\lambda}(\vec{r})=A^{(ij)}_{\lambda}(r)\, \hat{r}\!\:\!\otimes\!\:\!\hat{r} \:\!
+ B^{(ij)}_{\lambda}(r)\:\!(\Eins-\hat{r}\!\:\!\otimes\!\:\!\hat{r}) 
\label{eq:hij}%
\end{equation}
with $\lambda=\mathrm{s}$ for \ZT{self} and $\lambda=\mathrm{c}$ for \ZT{cross}, respectively, and 
the notation $r=\norm{\vec{r}}$ and $\hat{r}=\vec{r}/r$ for an arbitrary vector $\vec{r}$.
The self- and cross-interaction functions depend on the four mobility functions $A^{(ij)}_{\lambda}(r)$ and 
$B^{(ij)}_{\lambda}(r)$ with $\lambda\in\{\mathrm{s},\mathrm{c}\}$. 
With the \textit{method of reflections} \cite{HappelB1991,Dhont1996}, these mobility functions can be determined up to arbitrary order as an expansion 
in the inverse inter-particle distances. 
Up to fourth order, the mobility functions are given by\footnote{The identity 
$\Laplace_{\vec{r}^{(i)}_{k}}(\hat{r}^{(ij)}_{kl}\!\otimes\hat{r}^{(ij)}_{kl})
=(2\,\Eins-6\,\hat{r}^{(ij)}_{kl}\!\otimes\hat{r}^{(ij)}_{kl})/r^{(ij)2}_{kl}$ 
is useful for the derivation of Eqs.\ \eqref{eq:MFa}-\eqref{eq:MFd}.} 
{\allowdisplaybreaks
\begin{align}%
\begin{split}%
A^{(ij)}_{\mathrm{s}}(r)&=\mathcal{O}\Big(r^{-4}\Big) \,, 
\end{split}\label{eq:MFa}\\
\begin{split}%
B^{(ij)}_{\mathrm{s}}(r)&=\mathcal{O}\Big(r^{-4}\Big) \,, 
\end{split}\label{eq:MFb}\\
\begin{split}%
A^{(ij)}_{\mathrm{c}}(r)&=\frac{3}{2}\frac{R_{j}}{r} 
- \frac{1}{2}\frac{R^{2}_{i}R_{j}+R^{3}_{j}}{r^{3}} 
+ \mathcal{O}\Big(r^{-4}\Big) \,, 
\end{split}\label{eq:MFc}\\
\begin{split}%
B^{(ij)}_{\mathrm{c}}(r)&=\frac{3}{4}\frac{R_{j}}{r} 
+ \frac{1}{4}\frac{R^{2}_{i}R_{j}+R^{3}_{j}}{r^{3}} 
+ \mathcal{O}\Big(r^{-4}\Big) \,, 
\end{split}\label{eq:MFd}%
\end{align}}%
where $R_{i}$ denotes the radius of a colloidal particle of species $i$. 
Notice that Eqs.\ \eqref{eq:Dklij} and \eqref{eq:hij} together with the fourth-order approximations \eqref{eq:MFa}-\eqref{eq:MFd} 
of the mobility functions constitute a \textit{generalized Rotne-Prager approximation} for mixtures \cite{RotneP1969,Dhont1996}.   

The generalization of the derivation in Refs.\ \cite{RexL2008,RexL2009} leads to the following approximation of the 
diffusion tensor \eqref{eq:DTMa} for hydrodynamic pair-interactions:
\begin{equation}
\begin{split}%
D^{(ij)}_{\mathrm{HI}}(\vec{r},\rs\!,t)&= 
D^{(i)}_{0}\delta_{ij}\delta(\vec{r}-\rs)\big(\Eins\:\! c_{i}(\vec{r},t) + \mathrm{c}^{(i)}_{\mathrm{s}}(\vec{r},t)\big) \\
&\quad\:\!+ D^{(j)}_{0}\:\!\mathrm{h}^{(ij)}_{\mathrm{c}}(\vec{r}-\rs) \:\!c_{ij}(\vec{r},\rs\!,t) \;. 
\end{split}\raisetag{4ex}%
\end{equation}
Here, we introduced the functions
\begin{equation}
\mathrm{c}^{(i)}_{\mathrm{s}}(\vec{r},t) = \sum^{n}_{j=1}\int_{\R^{3}}\!\!\!\!\!\:\!\dif^{3}r'\, 
\:\!\mathrm{h}^{(ij)}_{\mathrm{s}}(\vec{r}-\rs) \:\!c_{ij}(\vec{r},\rs\!,t)
\end{equation}
and the two-particle concentrations    
\begin{equation}
c_{ij}(\vec{r},\rs\!,t) = \Tr(\rho(0)\hat{c}_{ij}(\vec{r},\rs\!,t)) 
\end{equation}
with the corresponding variables
\begin{equation}
\hat{c}_{ij}(\vec{r},\rs\!,t) = 
\sum^{N_{i}}_{k=1}\sum^{N_{j}}_{\begin{subarray}{c}l=1\\l\neq k\end{subarray}} 
\delta\big(\vec{r}-\vec{r}^{(i)}_{k}(t)\big) \delta\big(\rs-\vec{r}^{(j)}_{l}(t)\big) \,. 
\end{equation}
The two-particle variables $\hat{c}_{ij}(\vec{r},\rs\!,t)$ are assumed to relax much faster to local thermodynamic equilibrium 
than $\hat{c}_{i}(\vec{r},t)$ and $\hat{\varepsilon}(\vec{r},t)$.

\subsection{Approximation of the free-energy functional}
In order to determine the Helmholtz free-energy functional $\mathcal{F}[c_{1},\dotsc,c_{n},\varepsilon]$, 
which is needed in the EDDFT equations \eqref{eq:EDDFTa} and \eqref{eq:EDDFTb}, it is always possible to expand 
this functional with respect to $c_{i}(\vec{r},t)$, $\varepsilon(\vec{r},t)$, and their gradients 
taking general symmetry considerations into account \cite{PleinerB1996}. 

If the energy density can be neglected so that only an approximation for the functional $\mathcal{F}[c_{1},\dotsc,c_{n}]$ 
is needed, static density functional theory can be applied to derive such an approximation on a microscopic basis. 
The up to now most accurate approximation for $\mathcal{F}[c_{1},\dotsc,c_{n}]$ was derived in the framework of fundamental measure theory 
(see Ref.\ \cite{Roth2010} for a review).

\subsection{Special cases of the EDDFT equations}
The EDDFT equations \eqref{eq:EDDFTa} and \eqref{eq:EDDFTb} contain several special cases that are known from the literature or 
that are relevant for particular applications. Two of these special cases are addressed in this section. 
The first one is an isothermal binary mixture, where only two concentrations are present and the energy density can be neglected. 
As a second example, the hydrodynamic limit of the EDDFT equations is discussed.

\subsubsection{Isothermal binary mixture}
If the considered mixture consists only of $n=2$ different species of colloidal particles and the energy density can be assumed to be constant, 
the EDDFT equations \eqref{eq:EDDFTa} and \eqref{eq:EDDFTb} can be simplified to 
{\allowdisplaybreaks
\begin{align}%
\begin{split}%
&\dot{c}_{1}(\vec{r},t) = \beta D^{(1)}_{0} \Nabla_{\vec{r}}\!\cdot\! 
\big( \Eins\:\! c_{1}(\vec{r},t) + \mathrm{c}^{(1)}_{\mathrm{s}}(\vec{r},t) \!\:\!\big) 
\Nabla_{\vec{r}}\, c^{\natural}_{1}(\vec{r},t) \phantom{\int_{\R}} \\
&\;\; + \beta D^{(1)}_{0} \Nabla_{\vec{r}}\!\cdot\!\!\int_{\R^{3}}\!\!\!\!\!\:\!\dif^{3}r'\, 
\mathrm{h}^{(11)}_{\mathrm{c}}(\vec{r}-\rs) \:\!c_{11}(\vec{r},\rs\!,t)  
\Nabla_{\rs}c^{\natural}_{1}(\rs\!,t) \\
&\;\; + \beta D^{(2)}_{0} \Nabla_{\vec{r}}\!\cdot\!\!\int_{\R^{3}}\!\!\!\!\!\:\!\dif^{3}r'\, 
\mathrm{h}^{(12)}_{\mathrm{c}}(\vec{r}-\rs) \:\!c_{12}(\vec{r},\rs\!,t)  
\Nabla_{\rs}c^{\natural}_{2}(\rs\!,t) \;, 
\end{split}\label{eq:BMI}\raisetag{6em}\\[1ex]%
\begin{split}%
&\dot{c}_{2}(\vec{r},t) = \beta D^{(2)}_{0} \Nabla_{\vec{r}}\!\cdot\! 
\big( \Eins\:\! c_{2}(\vec{r},t) + \mathrm{c}^{(2)}_{\mathrm{s}}(\vec{r},t) \!\:\!\big) 
\Nabla_{\vec{r}}\, c^{\natural}_{2}(\vec{r},t) \phantom{\int_{\R}} \\
&\;\; + \beta D^{(1)}_{0} \Nabla_{\vec{r}}\!\cdot\!\!\int_{\R^{3}}\!\!\!\!\!\:\!\dif^{3}r'\, 
\mathrm{h}^{(21)}_{\mathrm{c}}(\vec{r}-\rs) \:\!c_{21}(\vec{r},\rs\!,t) 
\Nabla_{\rs}c^{\natural}_{1}(\rs\!,t) \\
&\;\; + \beta D^{(2)}_{0} \Nabla_{\vec{r}}\!\cdot\!\!\int_{\R^{3}}\!\!\!\!\!\:\!\dif^{3}r'\, 
\mathrm{h}^{(22)}_{\mathrm{c}}(\vec{r}-\rs) \:\!c_{22}(\vec{r},\rs\!,t)  
\Nabla_{\rs}c^{\natural}_{2}(\rs\!,t) \;. 
\end{split}\label{eq:BMII}\raisetag{6em}%
\end{align}}%
Here, the short-time diffusion coefficients \eqref{eq:Di} can be expressed by 
\begin{equation}
D^{(i)}_{0}=\frac{1}{\beta \:\! 6\pi\eta R_{i}} 
\end{equation}
with the dynamic (shear) viscosity $\eta$ of the molecular solvent.  
Hydrodynamic interactions between the colloidal particles are still taken into account by Eqs.\ \eqref{eq:EDDFTa} and \eqref{eq:EDDFTb}.
In the fourth-order approximation \eqref{eq:MFa}-\eqref{eq:MFd}, the functions $\mathrm{c}^{(i)}_{\mathrm{s}}(\vec{r},t)=\Null$ 
vanish ($\Null$ denotes the zero matrix) and the cross-interaction functions $\mathrm{h}^{(ij)}_{\mathrm{c}}(\vec{r})$ are 
\begin{equation}
\begin{split}%
\mathrm{h}^{(ij)}_{\mathrm{c}}(\vec{r}) &= \frac{3}{4}\frac{R_{j}}{\norm{\vec{r}}}  
\bigg( \Eins + \frac{\vec{r}\!\:\!\otimes\!\:\!\vec{r}}{\norm{\vec{r}}^{2}} \bigg) \\
&\quad\,+\frac{1}{4} \frac{R^{2}_{i}R_{j}+R^{3}_{j}}{\norm{\vec{r}}^{3}} 
\bigg( \Eins - 3\:\!\frac{\vec{r}\!\:\!\otimes\!\:\!\vec{r}}{\norm{\vec{r}}^{2}} \bigg) \,.  
\end{split}%
\end{equation}
As closure relations for the two-particle concentrations $c_{ij}(\vec{r},\rs\!,t)$ in the dynamical equations \eqref{eq:BMI}, 
the (exact) generalized Ornstein-Zernike equation for mixtures or simple analytical approximations that are known from the 
literature can be applied \cite{RexL2008,RexL2009}.

\subsubsection{The hydrodynamic limit}
The derived EDDFT equations \eqref{eq:EDDFTa} and \eqref{eq:EDDFTb} with the space- and time-dependent diffusion tensors 
\eqref{eq:DTMa}-\eqref{eq:DTMc} constitute an extension of the corresponding hydrodynamic equations to larger wave vectors $\vec{k}$ and 
frequencies $\omega$. In the hydrodynamic limit ($\vec{k}\to\vec{0}$, $\omega\to 0$), the EDDFT equations become  
{\allowdisplaybreaks%
\begin{align}%
\begin{split}%
\dot{c}_{i}(\vec{r},t) &= \sum^{n}_{j=1} \beta D^{(ij)}_{0}\!\Laplace_{\vec{r}}\,c^{\natural}_{j}(\vec{r},t) 
+\beta D^{(i\varepsilon)}_{0}\!\Laplace_{\vec{r}}\,\varepsilon^{\natural}(\vec{r},t) \;,
\end{split}\raisetag{5.2ex}\label{eq:EDDFTHLa}\\%
\begin{split}%
\dot{\varepsilon}(\vec{r},t) &= \sum^{n}_{j=1}\beta D^{(\varepsilon j)}_{0}\!\Laplace_{\vec{r}}\,c^{\natural}_{j}(\vec{r},t) 
+\beta D^{(\varepsilon\varepsilon)}_{0}\!\Laplace_{\vec{r}}\,\varepsilon^{\natural}(\vec{r},t) 
\end{split}\raisetag{2.5ex}\label{eq:EDDFTHLb}%
\end{align}}%
with the constant diffusion coefficients
{\allowdisplaybreaks%
\begin{align}%
\begin{split}%
D^{(ij)}_{0} &= \frac{1}{3} \int_{\R^{3}}\!\!\!\!\!\:\!\dif^{3}r 
\!\int^{\infty}_{0}\!\!\!\!\!\!\!\:\!\dif t\,\Tr\!\Big(\rho(0) \hat{\vec{J}}^{c_{i}}(\vec{r},t) \!\cdot\! 
\hat{\vec{J}}^{c_{j}}(\vec{0},0)\!\Big) \,,
\end{split}\label{eq:DTMHLa}\\%
\begin{split}%
D^{(i\varepsilon)}_{0} &= \frac{1}{3} \int_{\R^{3}}\!\!\!\!\!\:\!\dif^{3}r 
\!\int^{\infty}_{0}\!\!\!\!\!\!\!\:\!\dif t\,\Tr\!\Big(\rho(0) \hat{\vec{J}}^{c_{i}}(\vec{r},t) \!\cdot\! 
\hat{\vec{J}}^{\varepsilon}(\vec{0},0)\!\Big) \,,
\end{split}\label{eq:DTMHLb}\\%
\begin{split}%
D^{(\varepsilon\varepsilon)}_{0} &= \frac{1}{3} \int_{\R^{3}}\!\!\!\!\!\:\!\dif^{3}r 
\!\int^{\infty}_{0}\!\!\!\!\!\!\!\:\!\dif t\,\Tr\!\Big(\rho(0) \hat{\vec{J}}^{\varepsilon}(\vec{r},t) \!\cdot\! 
\hat{\vec{J}}^{\varepsilon}(\vec{0},0)\!\Big) 
\end{split}\label{eq:DTMHLc}%
\end{align}}%
and $D^{(\varepsilon i)}_{0}=D^{(i\varepsilon)}_{0}$. 
The hydrodynamic limit of the transport equations \eqref{eq:TGCij}-\eqref{eq:TGCee} for the 
time correlation functions \eqref{eq:DCij}-\eqref{eq:DCee} can be obtained analogously.

\subsection{Relation of EDDFT and MCT}
The MCT of glass transitions \cite{Goetze1991,Goetze2009} is a classical theory for the dynamics of liquids near the glass transition. 
Originally, MCT was constructed for the underdamped dynamics of atomic and molecular systems \cite{Goetze2009}, 
but it can also be derived for the overdamped dynamics of colloidal systems \cite{DieterichP1979,CichockiH1987,SzamelL1991}.
Like EDDFT, also MCT can be derived from the MZFT. This allows a comparison of these two theories on a common fundamental basis. 
In the following, we summarize the derivation of MCT and discuss its relation to DDFT and EDDFT.

\subsubsection{MCT for atomic and molecular systems}
The traditional form of MCT applies to a one-component system of equal spherical atoms or molecules of mass $m$. 
This system is characterized by a one-particle density field 
$\hat{c}(\vec{r},t)$ following the conservation law $\dot{\hat{c}}(\vec{r},t)+\Nabla_{\vec{r}}\!\cdot\!\hat{\vec{J}}^{\mathrm{c}}(\vec{r},t)=0$ 
with the density current $\hat{\vec{J}}^{\mathrm{c}}(\vec{r},t)$. 
In order to derive MCT, we switch to the Fou\-rier-La\-place space and utilize Eqs.\ \eqref{eq:CijFL}, where we omit the letters \ZT{eq} 
denoting equilibrium quantities and the tilde $\widetilde{\,}$ denoting quantities in the Fourier-Laplace space for reasons of clarity in this section. 
Near the glass transition, two variables are taken into account as relevant variables. 
These are the density field $\hat{c}(\vec{k},z)$ and the longitudinal component 
$\hat{j}^{\mathrm{L}}(\vec{k},z)=\vec{k}/k\cdot\hat{\vec{J}}^{\mathrm{c}}(\vec{k},z)$ of the density current $\hat{\vec{J}}^{\mathrm{c}}(\vec{k},z)$.
The transversal component of the density current, on the other hand, does not couple to density fluctuations and can therefore be neglected. 
While this was not the case in the context of EDDFT, here also the current associated with the density field has to be regarded as a relevant variable,
since there is no separation of time scales between these variables near the glass transition \cite{Goetze2009}.
We further define the concentration time autocorrelation function (dynamic structure factor) 
$C^{\mathrm{c}}(\vec{k},z)=\langle\Delta\hat{c}(\vec{k},z)|\Delta\hat{c}(\vec{k},0)\rangle$ and the current time autocorrelation function
$C^{\mathrm{j}}(\vec{k},z)=\langle\Delta\hat{j}^{\mathrm{L}}(\vec{k},z)|\Delta\hat{j}^{\mathrm{L}}(\vec{k},0)\rangle$ corresponding 
to the chosen relevant variables. 
With these definitions, application of Eqs.\ \eqref{eq:CijFL} leads directly to a dynamical equation for the  
normalized density time autocorrelation function $\phi^{\mathrm{c}}(\vec{k},z)=C^{\mathrm{c}}(\vec{k},z)/C^{\mathrm{c}}(\vec{k},0)$.
This dynamical equation is the \textit{MCT equation} \cite{Goetze2009}
\begin{equation}
\phi^{\mathrm{c}}(\vec{k},z) = \bigg(z+\frac{\Omega^{2}_{\mathrm{m}}(\vec{k})}{z-K^{\mathrm{j}}(\vec{k},z)}\bigg)^{-1}
\label{eq:MCTu}%
\end{equation}
with the frequency $\Omega_{\mathrm{m}}(\vec{k})$ that must not be confused with the (vanishing) frequency matrix in Eqs.\ \eqref{eq:CijFL}.
This frequency is given by $\Omega^{2}_{\mathrm{m}}(\vec{k})=C^{\mathrm{j}}(\vec{k},0)\vec{k}^{2}/C^{\mathrm{c}}(\vec{k},0)$
and $C^{\mathrm{j}}(\vec{k},0)=1/(\beta m)$. 
Furthermore, the current memory function $K^{\mathrm{j}}(\vec{k},z)$ in Eq.\ \eqref{eq:MCTu} is defined as 
$K^{\mathrm{j}}(\vec{k},z)=-\langle\Delta\dot{\hat{j}}^{\mathrm{L}}(\vec{k},0) 
|\hat{\mathcal{Q}}(z-\hat{\mathcal{L}}_{\hat{\mathcal{Q}}})^{-1}\hat{\mathcal{Q}}| 
\Delta\dot{\hat{j}}^{\mathrm{L}}(\vec{k},0)\rangle/C^{\mathrm{c}}(\vec{k},0)$.

\subsubsection{MCT for colloidal systems}
In case of a system of spherical colloidal particles that are suspended in a molecular solvent, 
a simpler MCT equation but with the same long-time behavior as Eq.\ \eqref{eq:MCTu} can be derived. 
This colloidal system is characterized by the short-time diffusion coefficient $D_{0}$ and concentration field $\hat{c}(\vec{k},z)$ 
of the colloidal particles. 
A similar derivation as before, but now with the appropriate Smoluchowski operator $\hat{\mathcal{L}}_{\mathrm{S}}$ instead of the 
Liouvillian $\hat{\mathcal{L}}$, leads to the \textit{MCT equation for colloidal systems} \cite{DieterichP1979,CichockiH1987,SzamelL1991}
\begin{equation}
\phi^{\mathrm{c}}(\vec{k},z) = \bigg(z+\frac{\Omega^{2}_{\mathrm{D}}(\vec{k})}{1-K^{\mathrm{j}}(\vec{k},z)}\bigg)^{-1}
\label{eq:MCTo}%
\end{equation}
with $\Omega^{2}_{\mathrm{D}}(\vec{k})=D_{0}\vec{k}^{2}/C^{\mathrm{c}}(\vec{k},0)$.
Notice that Eq.\ \eqref{eq:MCTo} is only of first order in $z$, while Eq.\ \eqref{eq:MCTu} is of second order.

\subsubsection{Comparison of EDDFT and MCT}
Although the derivation of MCT was only presented for the simple special case of a one-component system here, more general formulations of MCT exist  
that are like EDDFT, for example, also applicable to (colloidal) mixtures \cite{Goetze2009}. 
Even the incorporation of the energy density into MCT has already been discussed \cite{GoetzeL1989} in the literature. 
EDDFT and MCT are therefore two different general theories with overlapping fields of application. 
A possible relation of DDFT and MCT has been mentioned by Archer \cite{Archer2006,Archer2009}, but was not yet rigorously proven.
Archer showed that under certain approximations the traditional DDFT equation \cite{MarconiT1999,MarconiT2000,ArcherE2004} 
can be rearranged into a transport equation for the density time autocorrelation function, which matches the standard form \eqref{eq:MCTo} 
of MCT for colloidal systems.
However, his derivation, which suggests that MCT can be derived from DDFT, is not rigorous, since it involves a reinterpretation of 
the one-particle density field as a temporally coarse-grained density field.

\begin{figure}[ht]
\includegraphics[width=\linewidth]{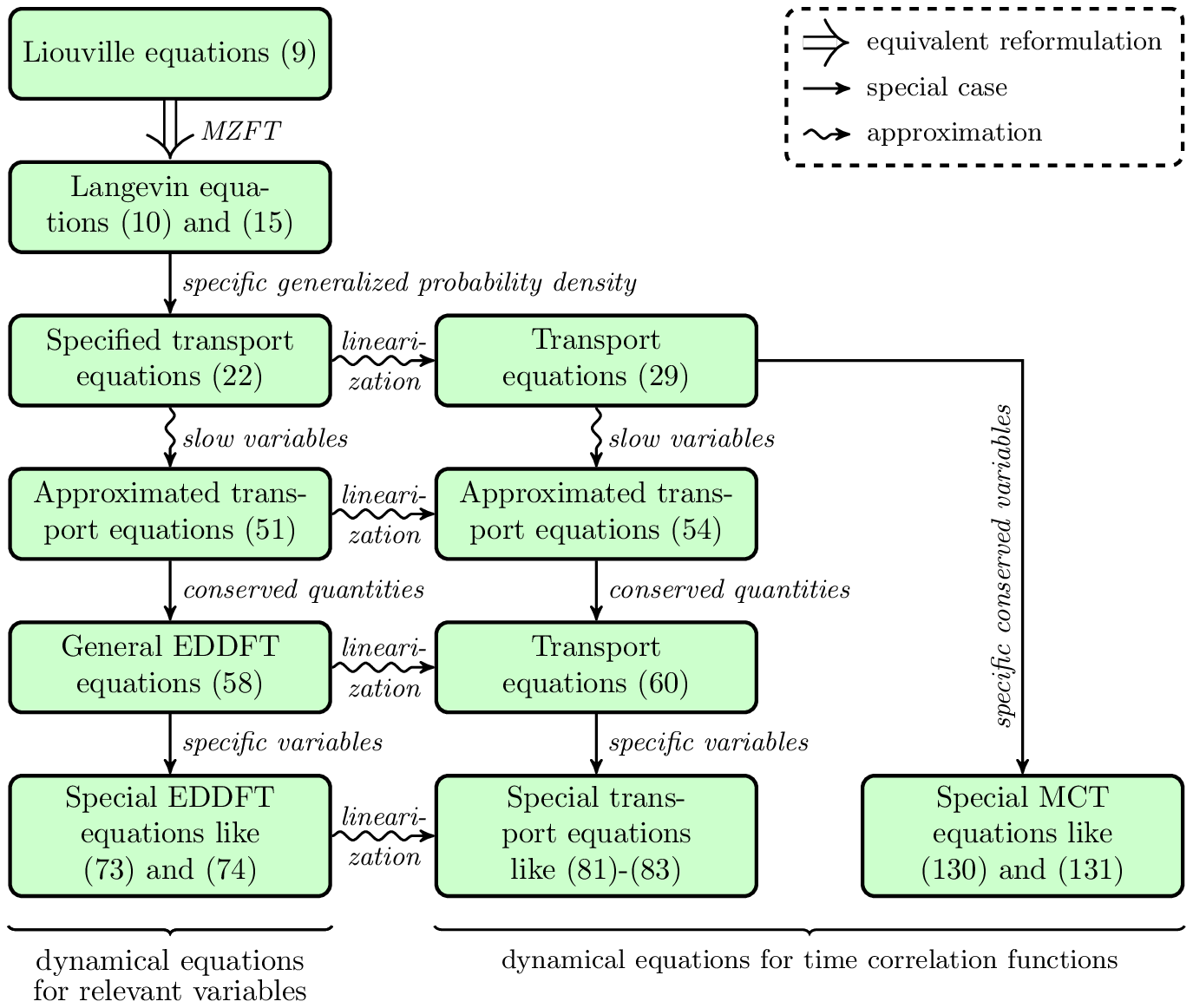}
\caption{\label{fig:DDFT}Illustration of the derivation of EDDFT and MCT using the MZFT.} 
\end{figure}

In contrast, our derivation of the EDDFT presented in this paper allows to compare both theories from a fundamental 
point of view. The derivation of EDDFT and MCT on the basis of the MZFT is illustrated in Fig.\ \ref{fig:DDFT}.
A comparison of the derivations of these theories makes clear that there are actually strong differences between EDDFT and MCT 
and that a rigorous derivation of one theory from the other is not possible.
An obvious and important difference between EDDFT and MCT results from the different approximations made in their derivations. 
While the derivation of EDDFT involves a Markovian approximation when slow variables are assumed (see Sec.\ \ref{sec:MA}), 
the MCT equations are non-Markovian -- a feature that becomes indeed relevant near the glass transition. 
A more detailed comparison reveals that EDDFT and MCT are rather complementary but not replaceable theories. 
While EDDFT has proven that it is successfully applicable to describe weakly correlated systems with low concentrations, it cannot be applied 
to the glassy dynamics of systems at very high densities, since the Markovian approximation in the derivation of the EDDFT equation  
can only be justified, if effects associated with long time tails can be neglected \cite{Grabert1982}. 
To the contrary, the derivation of MCT does not involve a Markovian approximation and has proven to be a useful analytical tool for the 
description of strongly correlated systems with high concentrations, where long time tails have to be taken into account \cite{Goetze2009}.
However, its derivation involves strong approximations, too, so that MCT fails when it is applied to 
weakly correlated dilute suspensions.

\section{\label{sec:conclu}Conclusions}
In this paper we have generalized classical dynamical density functional theory (DDFT) 
using the Mori-Zwanzig-Forster projection operator technique (MZFT) by
adding concentration fields and the energy density as variables. 
The resulting extended dynamical density functional theory (EDDFT) was compared to its hydrodynamic limit and to mode-coupling theory (MCT) 
revealing that EDDFT and MCT are complementary theories with different fields of application.
Our EDDFT framework shows that the MZFT is a flexible framework to incorporate thermal gradients 
(and other possible slow fields). 

We emphasize that, in principle, our EDDFT equations \eqref{eq:EDDFTa} and \eqref{eq:EDDFTb} treat concentration
and temperature gradients on arbitrary length scales even down to microscopic length scales of the average distance between the
colloidal particles.
The essential input for our EDDFT equations are functional derivatives, which can be obtained from equilibrium correlations, 
and diffusion tensors, which can be obtained from dynamical correlations.
An important challenge for the future is to apply this concept to actual temperature gradients in order to predict the Soret coefficient.

Guided by the application of the MZFT to various hydrodynamic systems including those 
with macroscopic degrees of freedom associated with spontaneously broken 
continuous symmetries \cite{Forster1974,BrandDG1979,BrandP1982b}, it will also be interesting to see to what extent one can generalize 
hydrodynamic considerations \cite{KadanoffM1963,HohenbergM1965} using correlation functions to larger wave vectors and frequencies.
 
A future generalization of the EDDFT equations should also take anisotropic colloidal particles 
with macroscopic degrees of freedom into account so that colloidal liquid crystals can be addressed \cite{WittkowskiL2011}. 
It will be important to compare such an approach to the results obtained previously for colloidal liquid crystals 
using a parametrization of the density with spherically symmetric, dipolar, and quadrupolar contributions 
\cite{Loewen2010,WittkowskiLB2010,WittkowskiLB2011,WittkowskiLB2011b}. 
Also the incorporation of the entropy density \cite{Schmidt2011} as a further variable would be an important task for the future.

Another challenge for the future is the potential use of the MZFT for systems driven far from thermodynamic equilibrium
for which a generalized thermodynamic potential \cite{GrahamH1971a,GrahamH1971b,Risken1972,Graham1978}
is not known. To address this question appears to be particularly important for active systems, 
which have increasingly come into focus over the last few years 
\cite{ParrishEK1999,KruseJJPS2005,MuhuriRR2007,GiomiML2008,BrandPS2011}. 

Recently, a similar approach using the MZFT for the one-particle density 
and the energy density was put forward by Espa\~{n}ol \cite{Espanol}.
This approach is based on an entropy functional formalism and provides explicit expressions for hard spheres.
However, mixtures and hydrodynamic interactions are not treated explicitly in this approach \cite{Espanol}.

\acknowledgments{We thank Pep Espa\~{n}ol for helpful discussions. 
This work was supported by the DFG within SPP 1296.}

\appendix
\section{\label{app:FLT}Integral transformations}
Since there are different definitions of the Fourier- and Laplace transformations in the literature, 
here we summarize the definitions that have been used in the context of the work presented.
In addition, two useful relations between the Fourier- and Laplace transformation are given.

\subsection{Fourier transformation}
The Fourier transformation of a space- and time-dependent function $X(\vec{r},t)$ is given by 
\begin{equation}
\begin{split}
\widetilde{X}(\vec{k},\omega) &= \int_{\R^{3}}\!\!\!\!\!\:\!\dif^{3}r \!\int_{\R}\!\!\hskip-0.5pt\dif t\, X(\vec{r},t) e^{\ii(\vec{k}\cdot\vec{r}-\omega t)} \;, \\
X(\vec{r},t) &= \frac{1}{(2\pi)^{4}} \!\int_{\R^{3}}\!\!\!\!\!\:\!\dif^{3}k\!\int_{\R}\!\!\hskip-0.5pt\dif \omega\,
\widetilde{X}(\vec{k},\omega) e^{-\ii(\vec{k}\cdot\vec{r}-\omega t)} \\
\end{split}
\end{equation}
with $\vec{k}\in\R^{3}$ and $\omega\in\R$.

\subsection{Laplace transformation}
The Laplace transformation of a time-dependent function $X(t)$ is given by 
\begin{equation}
\begin{split}
\widetilde{X}(z) &= \int^{\infty}_{0}\!\!\!\!\!\!\!\:\!\dif t\, X(t) e^{-zt} \;, \\
X(t) &= \frac{1}{2\pi\:\!\ii} \!\int^{c+\ii\infty}_{c-\ii\infty}\!\!\!\!\!\!\!\!\!\!\!\!\!\!\dif z\,
\widetilde{X}(z) e^{zt} \\
\end{split}
\end{equation}
with $z\in\C$ and the real part $\Re(z)>0$. The expression for the inverse Laplace transformation is known as Bromwich integral and contains a 
constant $c>z_{0}$, where $z_{0}$ is the convergence abscissa of $\widetilde{X}(z)$.

\subsection{Useful relations}
The Fourier transformed function $\widetilde{X}(\omega)$ and the Laplace transformed function $\widetilde{X}(z)$ can directly be transformed 
into each other. 
With the residue theorem, the following map from $\widetilde{X}(\omega)$ to $\widetilde{X}(z)$ can be proven:
\begin{equation}
\widetilde{X}(z)=\frac{\ii}{2\pi}\!\int_{\R}\!\!\hskip-0.5pt\dif \omega\,\frac{\widetilde{X}(\omega)}{\omega+\ii z} \;.
\end{equation}
A complementary map from $\widetilde{X}(z)$ to $\widetilde{X}(\omega)$ is given by 
\begin{equation}
\widetilde{X}(\omega)=\limeps \!\Big( \widetilde{X}(z)\big\rvert_{z=\ii\omega+\epsilon} 
-\widetilde{X}(z)\big\rvert_{z=\ii\omega-\epsilon} \Big) \,.
\end{equation}
It follows directly from the (special) \textit{Sokhotski-Plemelj theorem} \cite{Forster1990,Greiner2009}
\begin{equation}
\limeps \frac{1}{x\mp\ii\:\!\epsilon} = \PP\,\,\frac{1}{x} \pm \ii\:\!\pi\:\! \delta(x) \;,
\end{equation}
where $\operatorname{P}$ denotes the Cauchy principal value.

\bibliography{References}
\end{document}